\documentclass[aps,prd,onecolumn,preprintnumbers,nofootinbib]{revtex4}
\pdfoutput=1
\usepackage{natbib}

\preprint{SACLAY-T09-054}
\preprint{NORDITA-2009-29}
\usepackage{bm}
\usepackage{graphicx}
\usepackage{color}
\usepackage{amsmath}

\newcommand{\beq}{\begin{equation}}
\newcommand{\eeq}{\end{equation}}
\newcommand{\bdm}{\begin{displaymath}}
\newcommand{\edm}{\end{displaymath}}
\newcommand{\bea}{\begin{eqnarray}}
\newcommand{\eea}{\end{eqnarray}}
\newcommand{\bt}{\begin{tabular}}
\newcommand{\et}{\end{tabular}}
\newcommand{\xv}{{\bf x}}
\newcommand{\kv}{{\bf k}}

\newcommand{\qv}{{\bf q}}
\newcommand{\pv}{{\bf p}}

\newcommand{\intq}{\int\!\!d^3 q}

\newcommand{\tW}{\widetilde{W}_R}

\def\d{\delta}

\def\Mpc{\, h^{-1} \, {\rm Mpc}}

\def\kMpc{\, h \, {\rm Mpc}^{-1}}

\def\fNL{f_{NL}}
\def\gNL{g_{NL}}

\def\fNLl{f_{NL}^{\rm loc.}}

\def\fNLe{f_{NL}^{\rm eq.}}

\def\la{\langle}
\def\ra{\rangle}
\def\O{{\mathcal O}}

\def\la{\langle}
\def\ra{\rangle}

\begin{document}

\title{1-loop Perturbative Corrections to the Matter and Galaxy Bispectrum with non-Gaussian Initial Conditions}

\author{Emiliano Sefusatti}

\affiliation{Institut de Physique Th\'eorique, CEA, IPhT, F-91191 Gif-sur-Yvette, France}
\affiliation{Nordita, Nordic Institute for Theoretical Physics, 106 91 Stockholm, Sweden}
\email{emiliano.sefusatti@cea.fr}

\begin{abstract}
We derive the expressions for the 1-loop corrections in cosmological, Eulerian, perturbation theory to the {\it matter bispectrum} and to the {\it galaxy bispectrum}, assuming local galaxy bias, in presence of non-Gaussian initial conditions. We compute them explicitly for the particular case of non-vanishing initial bispectrum and trispectrum in the {\it local} model and for a non-vanishing initial bispectrum alone for the {\it equilateral} model of primordial non-Gaussianity. While the primordial contribution to the {\it matter} bispectrum for values compatible with CMB observations is dominant over the component due to gravitational instability at large scales, 1-loop perturbative corrections due to non-Gaussian initial conditions correspond to just a few percent of the gravity-induced bispectrum at mildly non-linear scales, similarly to what happens for the matter power spectrum. However, in the perturbative expansion for the {\it galaxy} bispectrum, 1-loop diagrams arising from non-linear bias are responsible for significant large-scale contributions, indeed exceeding the primordial component, {\it both} for the local and equilateral model. We study the peculiar dependence on scale and on the shape of the triangular configurations of such additional terms, similar in their origin to the large-scale corrections to the halo and galaxy power spectra that raised significant interest in the recent literature. 
\end{abstract}


\maketitle

\section{Introduction}

Recent observations of the Cosmic Microwave Background (CMB) have confirmed that primordial perturbations in the matter density are well described by a Gaussian random field, whose statistical properties are encoded in its two-point correlation function. Constraints on higher-order correlations such as the three-point function or its Fourier transform, the bispectrum, are model-dependent but generically consistent with a level of non-Gaussianity below $0.1\%$, \citep{KomatsuEtal2009B,SmithSenatoreZaldarriaga2009}.  Indeed, the simplest models of inflation typically predict departures from Gaussianity a few orders of magnitude smaller then current limits, \citep{AcquavivaEtal2003,Maldacena2003}. It is therefore evident that a possible detection in future observations of a significantly larger non-Gaussian component would rule-out a single-field slow-roll model for inflation and greatly improve our ability to discriminate among the currently wide range of available phenomenological models of the early Universe (for a review see, for instance, \citep{BartoloEtal2004}). 

Measurements of the bispectrum of the CMB temperature fluctuations have been, so far, the most powerful probe of primordial non-Gaussianity. On the other hand, the effects of non-Gaussian initial conditions on the growth of matter perturbations and on large-scale structure observables have been studied for quite a long time, focusing, in particular, on corrections to the high-mass tail of the cluster abundance \citep{LucchinMatarrese1988,ColafrancescoLucchinMatarrese1989,ChiuOstrikerStrauss1998,RobinsonGawiserSilk2000,MatarreseVerdeJimenez2000,RobinsonBaker2000,SefusattiEtal2007,LoVerdeEtal2008,KamionkowskiVerdeJimenez2009,MaggioreRiotto2009C} and to cumulants \citep{FryScherrer1994,ChodorowskiBouchet1996,DurrerEtal2000} and higher-order correlation functions \citep{Scoccimarro2000A,VerdeEtal2000,ScoccimarroSefusattiZaldarriaga2004,SefusattiKomatsu2007} of the galaxy distribution. Until recently, the galaxy bispectrum has been considered the most promising observable, expected to provide comparable or even better constraints on non-Gaussianity from measurements in future high-redshift, large-volume galaxy surveys, \citep{SefusattiKomatsu2007,SefusattiEtal2009}.

This picture radically changed during the last year after \citet{DalalEtal2008} showed, in numerical simulations, that a relatively small departure from Gaussianity can  have a large effect on the power spectrum of dark matter halos. These results, somehow anticipated by early theoretical works on the correlators of the distribution of peaks in the matter density field \citep{GrinsteinWise1986,MatarreseLucchinBonometto1986},  attracted significant attention  \citep{SlosarEtal2008,Seljak2009,Slosar2009,MatarreseVerde2008,CarboneVerdeMatarrese2008,AfshordiTolley2008,McDonald2008,TaruyaKoyamaMatsubara2008,DesjacquesSeljakIliev2009,PillepichPorcianiHahn2008,GrossiEtal2009,JeongKomatsu2009,FedeliMoscardiniMatarrese2009,Oguri2009,Valageas2009}, finding further confirmation in independent numerical results, \citep{DesjacquesSeljakIliev2009,PillepichPorcianiHahn2008,GrossiEtal2009}. Most importantly, they allowed to derive constraints on the amplitude of primordial non-Gaussianity of the order of CMB limits from {\it current} large-scale structure observations, \citep{SlosarEtal2008,AfshordiTolley2008}.

These first works focused their attention on the effects on the {\it power spectrum} of biased objects, halos and galaxies, proposing quite different theoretical descriptions. For instance, starting from the expression defining the specific non-Gaussian model assumed for the N-body simulations of \citet{DalalEtal2008}, a correction to the halo bias can be derived in the framework of the peak-background split, \citep{DalalEtal2008,SlosarEtal2008,AfshordiTolley2008}. This approach, however, seems to be hardly extensible to different models of primordial non-Gaussianity. An alternative description, proposed by \citet{MatarreseVerde2008} and based on the statistics of high-peaks in the matter density, provides a correction to the power spectrum of the peak distribution in term of the generic primordial matter bispectrum and can be therefore applied to any non-Gaussian model. A similar result, but derived in terms of 1-loop perturbative corrections to the galaxy power spectrum under the assumption of local, Eulerian bias in position space, has been suggested by \citet{TaruyaKoyamaMatsubara2008}. Finally an even different interpretation has been considered by \citet{McDonald2008}.

While all the proposed corrections to the linear bias function of the halo power spectrum present the same functional dependence on redshift and scale, which is quite generic when non-Gaussian initial conditions are considered, they also make distinct and relevant approximations. Despite the fact that a satisfactory agreement with the halo power spectrum measured in numerical simulations has been shown by \citet{GrossiEtal2009} when a correction factor, possibly accounting for ellipsoidal collapse of virialized objects, is included, we believe that more work is still required to provide a proper theoretical model for this effect applicable on a large range both in mass and scales {\it and} to a generic model of primordial of non-Gaussianity. 

A natural test of the different approaches mentioned above consists, in this respect, in the derivation of predictions for higher-order correlation functions, and, in the first place, for the {\it bispectrum}. A first step in this direction has been made by \citet{JeongKomatsu2009}, which applied to the bispectrum the theoretical description of \citet{MatarreseVerde2008} based on the distribution of high density peaks. In this paper we pursue the same goal extending instead the results of \citet{TaruyaKoyamaMatsubara2008} for the matter and galaxy power spectra to the matter and galaxy bispectra. For the galaxy bispectrum, we therefore assume a local bias prescription and compute all tree-level and 1-loop contributions in Eulerian Perturbation Theory up to a given order in the linear matter density field. We will show that some of these contributions, including a term corresponding to the one considered by \citet{JeongKomatsu2009}, are indeed dominant at large scales, in a similar way as in the power spectrum case. We will show, moreover, that, unlike the power spectrum, the galaxy bispectrum receives significant corrections also for an equilateral model of non-Gaussianity. 

It is reasonable to assume, based on our results, that previous predictions for the constraints on primordial non-Gaussianity from measurements of the galaxy bispectrum in current and future redshift surveys, \citep{ScoccimarroSefusattiZaldarriaga2004,SefusattiKomatsu2007} are likely to improve significantly. In fact, we can conservatively expect that for a non-Gaussian parameter, as for other cosmological parameters, the bispectrum should be able to provide at least comparable results as the ones provided by power spectrum measurements, \citep{SefusattiScoccimarro2005,SefusattiEtal2006}.
A proper assessment of the expected constraints on non-Gaussianity, however, will have to wait future work, and, in particular, necessary comparisons with N-body simulations. It is worth reminding that our assumption of a local bias does not include a prediction for the corresponding bias parameters, which are usually derived, for Gaussian initial conditions, in the framework of the Halo Model. However, it should provide the functional form of the most relevant contributions to the galaxy bispectrum which is the main result of this paper.

In Section~\ref{sec:PNG} we discuss relevant inflationary models responsible for a possibly observable non-Gaussian component in the density perturbations. We introduce in particular the local and equilateral expressions for the bispectrum of the initial curvature perturbations that will be used as examples in the rest of the paper. In Section~\ref{sec:matterPT} we summarize the Perturbation Theory (PT) approach to gravitational clustering of the matter density field and review previous results on the matter power spectrum in presence of non-Gaussian initial conditions. We then derive the 1-loop corrections to the matter bispectrum and the tree-level contributions to the matter trispectrum, {\it i.e.} the four-point function in Fourier space. In Section~\ref{sec:galaxyPT} we consider the perturbative expressions for the correlators of the galaxy density field under the assumption of local bias. In particular we derive the 1-loop corrections to the galaxy bispectrum and compute them for the local and equilateral models of primordial non-Gaussianity. In Section~\ref{sec:peaks} we discuss similarities and differences of our results with those recently proposed by \citet{JeongKomatsu2009}. Finally, we present our conclusions in Section~\ref{sec:conclusions}.

\section{Models of Primordial non-Gaussianity}
\label{sec:PNG}

Initial conditions for the evolution of matter perturbations are usually given in terms of the early-times correlators of the matter density field, $\delta$, or of the curvature fluctuations, $\Phi$. For Gaussian initial conditions, all statistical properties of the initial fields are encoded in their two-point function, or the power spectrum, in Fourier space. Small departures from Gaussianity can be characterized by non-vanishing three- and four-point functions, or their Fourier transform, respectively, the bispectrum and trispectrum. In the following sections we will study the effects on the late-times matter and galaxy bispectrum due to two common phenomenological models for non-Gaussian, primordial curvature perturbations, the {\it local} model \citep{SalopekBond1990,SalopekBond1991,GanguiEtal1994,VerdeEtal2000,KomatsuSpergel2001}, and the {\it equilateral} model, \citep{BabichCreminelliZaldarriaga2004}. 

The local model describes a small correction to Gaussian initial conditions, represented by the expression for Bardeen's curvature perturbations $\Phi$, in position space
\beq
\label{eq:phiNG}
\Phi(\xv) = \phi(\xv) + \fNL \left[ \phi^2(\xv) - \langle \phi^2(\xv) \rangle\right]+\gNL\phi^3(\xv),
\eeq
where $\phi$ represents the Gaussian component and the second and third term on the r.h.s. the quadratic and cubic, non-Gaussian components. We assume here the CMB convention for the definition of the $\fNL$ parameter, which implies that the curvature $\Phi$ is evaluated during matter domination. The local model typically describes non-linearities in the relation between the inflaton and curvature perturbations~\cite{SalopekBond1990,SalopekBond1991,GanguiEtal1994}, and in general models where non-Gaussianity is produced outside the horizon such as curvaton models~\cite{LythUngarelliWands2003}, inhomogeneous reheating \citep{DvaliGruzinovZaldarriaga2004A,DvaliGruzinovZaldarriaga2004B} or multiple field inflation \citep{BernardeauUzan2002}.

From Eq.~(\ref{eq:phiNG}) one can derive the following expression for the leading contribution to the curvature bispectrum,
\beq
\label{eq:Blc}
B_\Phi^{\rm loc.}(k_1, k_2, k_3) = 2 \fNL P_\Phi(k_1)P_\Phi(k_2)+ {\rm 2~perm.},
\eeq  
with the curvature power spectrum $P_\Phi(k)$ defined in terms of the Gaussian component alone as $\langle\phi({\bf k}_1) \phi({\bf k}_2) \rangle = \delta_D^{(3)}(\kv_{12}) P_\Phi(k_1)$, where we introduce the notation $\kv_{ij}\equiv\kv_i+\kv_j$. Similarly, the curvature trispectrum will be given by,
\bea
\label{eq:Tlc}
T_\Phi^{\rm loc.}(\kv_1,\kv_2,\kv_3,\kv_4) & = & 4 \fNL^2 P_\Phi(k_1)P_\Phi(k_2)\left[P_\Phi(k_{13})+P_\Phi(k_{14})\right]+ {\rm 5~perm.}\nonumber\\
& & +~6\gNL P_\Phi(k_1)P_\Phi(k_2)P_\Phi(k_3)+{\rm 3~perm.}
\eea  
Notice that this quantity depends on six variables, which can be chosen to be the magnitudes of the four wavenumbers plus the magnitudes of two sums like, for instance, $k_{12}$ and $k_{13}$. 

The factor $\fNL^2$ in front of the first term on the l.h.s. of Eq.~(\ref{eq:Tlc}) is clearly a consequence of the definition of Eq.~(\ref{eq:phiNG}). However, inflationary models predicting an initial bispectrum corresponding to Eq.~(\ref{eq:Blc}), can generically predict an initial trispectrum of the form of Eq.~(\ref{eq:Tlc}) but with a different normalization factor, sometimes denoted as $\tau_{NL}$ (see, for instance, \citep{BernardeauBrunier2007}). Such distinct quantity typically presents a  dependence on the parameters of the inflationary model different than $\fNL$. While we will consider, for simplicity, Eq.~(\ref{eq:Tlc}) as our model for the initial local trispectrum, we should keep in mind that the relation between the amplitudes of $B_\Phi$ and $T_\Phi$ is, in general, less trivial.

Current limits on the value of $\fNLl$ from measurements of the CMB bispectrum correspond to $-4<\fNLl<80$ ($95\%$ C.L.), \citep{SmithSenatoreZaldarriaga2009}, while large-scale structure constraints, exploiting the effect on the power spectrum of highly biased objects, are given by $-31<\fNLl<70$ ($95\%$ C.L.), \citep{SlosarEtal2008}, although somehow different analyses provide higher values, \citep{AfshordiTolley2008}. 

As a complementary expression to the bispectrum of the local model, \citet{BabichCreminelliZaldarriaga2004} introduced an {\it equilateral} model of primordial non-Gaussianity defined by the bispectrum
\bea
\label{eq:Beq}
B_{\Phi}^{\rm eq.}(k_1, k_2, k_3) & = &  
6\fNL\left[- P_\Phi(k_1) P_\Phi(k_2) + {\rm 2~perm.}
       - 2 P^{2/3}_\Phi(k_1)P^{2/3}_\Phi(k_2)P^{2/3}_\Phi(k_3) \right.
\nonumber\\
& &    \left. + P^{1/3}_\Phi(k_1)P^{2/3}_\Phi(k_2)P_\Phi(k_3) + {\rm 5~perm.}\right].
\eea
This functional form assumes larger values for equilateral triangular configurations, that is when $k_1\simeq k_2\simeq k_3$, while the local model is more sensitive to squeezed configurations where $k_1\simeq k_2\ll k_3$. The normalization of the expression in Eq.~(\ref{eq:Beq}) has been defined in a such a way as to provide an identical value as the local expression, Eq.~(\ref{eq:Blc}) for equilateral configurations, $k_1=k_2=k_3$ when $\fNLl=\fNLe$. It has been shown, \cite{BabichCreminelliZaldarriaga2004,CreminelliEtal2006,FergussonShellard2008}, that the equilateral model for the bispectrum closely approximates the bispectrum predicted by inflationary models with non-canonical kinetic terms like DBI inflation \citep{AlishahihaEtal2004}, Ghost inflation \citep{ArkaniHamedEtal2004} or higher derivatives \citep{Creminelli2003,ChenEtal2007}. Current limits on the amplitude of equilateral non-Gaussianity are $-151<\fNLe<253$ ($95\%$ C.L.), \citep{KomatsuEtal2009B}. 

While we limit ourselves to these two models of primordial non-Gaussianity, it should be stressed that several other functional forms for the initial bispectrum with peculiar dependencies on the shape of the triangular configurations are predicted by inflationary models in the literature (see \citep{FergussonShellard2008} and references therein).

The matter overdensity in Fourier space $\d_\kv$ is related to the curvature perturbations $\Phi_\kv$ by Poisson's equation as
\beq
\d_\kv(z)=M(k,z)~\Phi_\kv,
\eeq
where we introduce the function
\beq
M(k,z)=\frac{2}{3}\frac{k^2T(k)D(z)}{\Omega_mH_0^2},
\eeq
with $T(k)$ being the matter transfer function and $D(z)$ the growth factor. Notice that since we conform to the CMB convention for the definition of the non-Gaussian parameter $\fNL$, we denote with $\Phi$ the {\it primordial} curvature perturbations, {\it i.e.} evaluated during the matter dominated era, not their value linearly extrapolated at present time.

The initial correlators of the matter density field are related to the correlators of the curvature perturbations, in general as 
\beq
\la\d_{\kv_1}\cdots\d_{\kv_n}\ra=M(k_1,z)\cdots M(k_n,z)\la\Phi_{\kv_1}\cdots\Phi_{\kv_n}\ra,
\eeq
so that, in particular, the linear power spectrum is given by 
\beq
P_0(k)=M^2(k,z)P_\Phi(k),
\eeq
while the initial bispectrum and trispectrum are given respectively by
\beq\label{eq:B0}
B_0(k_1,k_2,k_3)=M(k_1)M(k_2)M(k_3)B_\Phi(k_1,k_2,k_3),
\eeq
and
\beq\label{eq:T0}
T_0(\kv_1,\kv_2,\kv_3,\kv_4)=M(k_1)M(k_2)M(k_3)M(k_4)T_\Phi(\kv_1,\kv_2,\kv_3,\kv_4).
\eeq

In the following sections we will study the effects of the local and equilateral models of primordial non-Gaussianity on the evolved matter bispectrum and trispectrum in the framework of Perturbation Theory and, with a similar perturbative approach based on the assumption of local galaxy bias, on the galaxy bispectrum. 

\section{Matter correlators and non-Gaussian initial conditions}
\label{sec:matterPT}

In Fourier space, the perturbative solution to the equations of gravitational instability for the matter overdensity can be formally expressed as \citep{BernardeauEtal2002},
\beq\label{eq:PT}
\d_\kv=\d_\kv^{(1)}+\d_\kv^{(2)}+\d_\kv^{(3)}+\ldots,
\eeq
where
\beq\label{eq:PTterm}
\d_\kv^{(n)}\equiv\int d^3q_1\ldots d^3q_nF_n(\qv_1,\ldots,\qv_n)\d_{\qv_1}^{(1)}\ldots\d_{\qv_n}^{(1)},
\eeq
where we keep implicit the time dependence as $\d_{\qv}^{(1)}\sim D(z)$, $D(z)$ being the linear growth factor, and where $F_n(\qv_1,...,\qv_n)$ is the symmetrized kernel for the $n$-th order solution.

As shown explicitly in \citep{CrocceScoccimarro2006A}, the expansion of Eq.~(\ref{eq:PT}) leads to an ill-defined perturbative solution for the power spectrum, characterized by large cancellations between contributions of the same order. New approaches to the physics of gravitational instability have been introduced with the Renormalized Perturbation Theory in \citep{CrocceScoccimarro2006A,CrocceScoccimarro2006B,CrocceScoccimarro2008}, and 
with the Renormalization Group formalism in \citep{MatarresePietroni2007} leading to a well behaved perturbative expansion of the matter power spectrum and more accurate predictions of simulations results. A first step in extending these techniques to higher-order correlators can be found in \citep{BernardeauCrocceScoccimarro2008}. We will leave the extension of renormalized perturbation theory or renormalization group approaches to non-Gaussian initial conditions to future work (but see \citet{IzumiSoda2007}), limiting ourselves to observe that traditional perturbation theory can still provide a satisfactory description of the impact of non-Gaussianities on the matter bispectrum, particularly since, as we will show, this turns-out to be negligible with respect to the Gaussian component. We also notice that, with this work, we are filling a gap in the literature that could serve as a starting point for future investigations. 

For all numerical evaluations in this Section, as well as in the following, we assume a flat $\Lambda$CDM cosmology with the WMAP 5-yr, \citep{KomatsuEtal2009B}, cosmological parameter values $\Omega_b=0.044, \Omega_c=0.214, H_0=71.9, n_s=0.96, n_t=0.0, \tau=0.087$ and with a r.m.s. of the matter fluctuations corresponding to $\sigma_8=0.8$.

\subsection{The matter power spectrum}

In this section we summarize, for completeness and because they will be useful in the following sections, previous results in cosmological perturbation theory relative to the matter power spectrum. Specifically, 1-loop perturbative corrections to the matter power spectrum have been studied in \citep{MakinoSasakiSuto1992,JainBertschinger1994} for Gaussian initial conditions. Additional contributions up to the same order in PT for generic non-Gaussian initial conditions have been derived in \citep{TaruyaKoyamaMatsubara2008} and computed for the local and equilateral non-Gaussian models assuming a non-vanishing initial bispectrum. Notice that, while in the Gaussian case all 1-loop corrections are of the same order, ${\mathcal O(\d_0^4)}$, in terms of the linear matter overdensity $\d_0$, for non-Gaussian initial conditions, we have an extra term---denoted as $P_{12}$ in \citep{TaruyaKoyamaMatsubara2008}---of order ${\mathcal O(\d_0^3)}$. Similarly, 2-loop corrections consist of contributions of several different powers in the initial field $\d_0$. We choose to write the perturbative solution for the matter power spectrum as a series of powers of the initial field $\d_0$, indicating each contribution of order $\O({\d_0^p})$ as $P^{(p)}$\footnote{This is different from the more common notation where $P^{(p)}$ indicates the $p$-loop correction.}, so that 
\beq
\label{eq:Pexp}
P(k)=P^{(2)}(k)+P^{(3)}(k)+P^{(4)}(k)+\O({\d_0^5}),
\eeq 
where, for Gaussian initial conditions, all terms with odd index vanish. Introducing the notation 
\beq
\d_D(\kv_{12})P_{ij}(k_1)\equiv\la\d_{\kv_1}^{(i)}\d_{\kv_2}^{(j)}\ra\,+\,{\rm cyc.},
\eeq
we have $P^{(2)}= P_{11}\equiv P_0$ being the linear matter power spectrum, while 
\bea
P^{(3)}(k)
& = & P_{12},
\\
P^{(4)}(k)
& = & 
P_{22}^I+P_{22}^{II}+P_{13}^I+P_{13}^{II},
\eea
with
\bea
P_{12}
& = &
2\intq F_2(\qv,\kv-\qv)~B_0(k,q,|\kv-\qv|),
\\
P_{22}^I
& = & 
2\intq F_2^2(\qv,\kv-\qv)~P_0(q)~P_0(|\kv-\qv|),
\\
P_{22}^{II}
& = & 
\intq~d^3 p F_2(\pv,\kv-\pv)F_2(\qv,-\kv-\qv)~T_0(\pv,\kv-\pv,\qv,-\kv-\qv),
\\
P_{13}^I
& = & 
6~P_0(k)\intq F_3(\kv,\qv,-\qv)~P_0(q),
\\
P_{13}^{II}
& = & 
2\intq~d^3pF_3(\pv,\qv,\kv-\pv-\qv)~T_0(-\kv,\pv,\qv,\kv-\pv-\qv).
\eea
In the equations above $B_0$ and $T_0$ represent the initial bispectrum and trispectrum, respectively, related to the higher-order correlators of the curvature perturbations according to Eq.~(\ref{eq:B0}) and Eq.~(\ref{eq:T0}). $P_{12}$ is the 1-loop non-Gaussian contributions studied by \citep{TaruyaKoyamaMatsubara2008}, the terms $P_{22}^I$ and $P_{13}^{I}$ constitute the Gaussian 1-loop corrections, \citep{MakinoSasakiSuto1992,JainBertschinger1994}, while $P_{22}^{II}$ and $P_{13}^{II}$ are 2-loop corrections vanishing for Gaussian initial conditions.
We will consistently ignore, for simplicity, in our evaluations throughout the paper all 2-loop contributions.

\begin{figure}[t]
\begin{center}
{\includegraphics[width=1\textwidth]{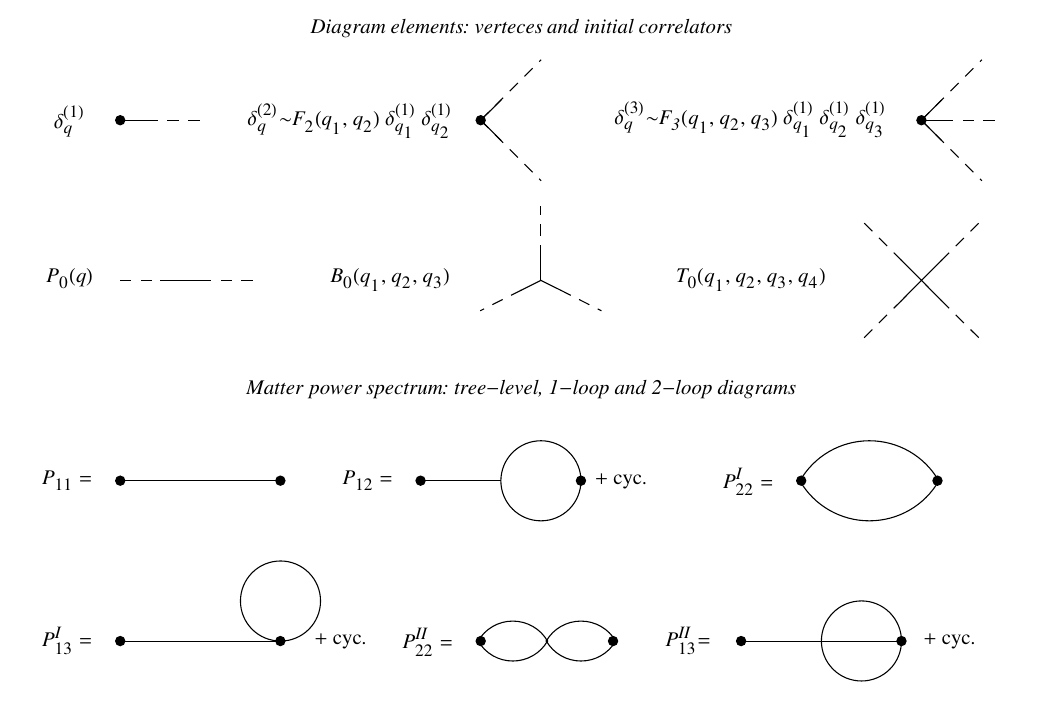}}
\caption{Elements of the diagrammatic representation of perturbative corrections of matter correlators and diagrams representing the tree-level ($P_{11}$), 1-loop ($P_{12}$, $P_{22}^I$ and $P_{13}^I$) and 2-loop ($P_{22}^{II}$ and $P_{13}^{II}$) contributions to the matter power spectrum up to terms of order $\O(\d_0^5)$. Lines connecting two points correspond to the linear power spectrum $P_0(k)$,  the three connected lines in the $P_{12}$ term corresponds to the initial bispectrum $B_0$ and the four connected lines in $P_{22}^{II}$ and $P_{13}^{II}$ correspond to the initial trispectrum $T_0$.}
\label{fig:graphsPm}
\end{center}
\end{figure}
The contributions to the solution for matter correlators in cosmological perturbation theory can find a representation in terms of diagrams (see \citep{BernardeauEtal2002}), often helpful to identify different terms and their characteristics. In Fig.~\ref{fig:graphsPm} we show the diagrams representing the tree-level, 1-loop and 2-loop contributions to the matter power spectrum in the expressions above. The upper half shows, in particular, the elements entering the diagrammatic representation, given by verteces corresponding to the perturbative solution of Eq.~(\ref{eq:PT}) and the initial correlators joining them. A simple line connecting two points corresponds to the linear power spectrum $P_0(k)$ while the three connected lines---as those in the $P_{12}$ term---correspond to the initial bispectrum $B_0$ and the four connected lines---as in $P_{22}^{II}$ and $P_{13}^{II}$---correspond to the initial trispectrum $T_0$. It is easy to see how non-vanishing higher-order correlators such as $B_0$ and $T_0$ provide several extra contributions to the non-linear matter power spectrum with respect to those present for Gaussian initial conditions, corresponding to a larger variety of possible diagrams. This will be even more evident for the non-linear bispectrum.

\subsection{The matter bispectrum}

\begin{figure}[t]
\begin{center}
{\includegraphics[width=1\textwidth]{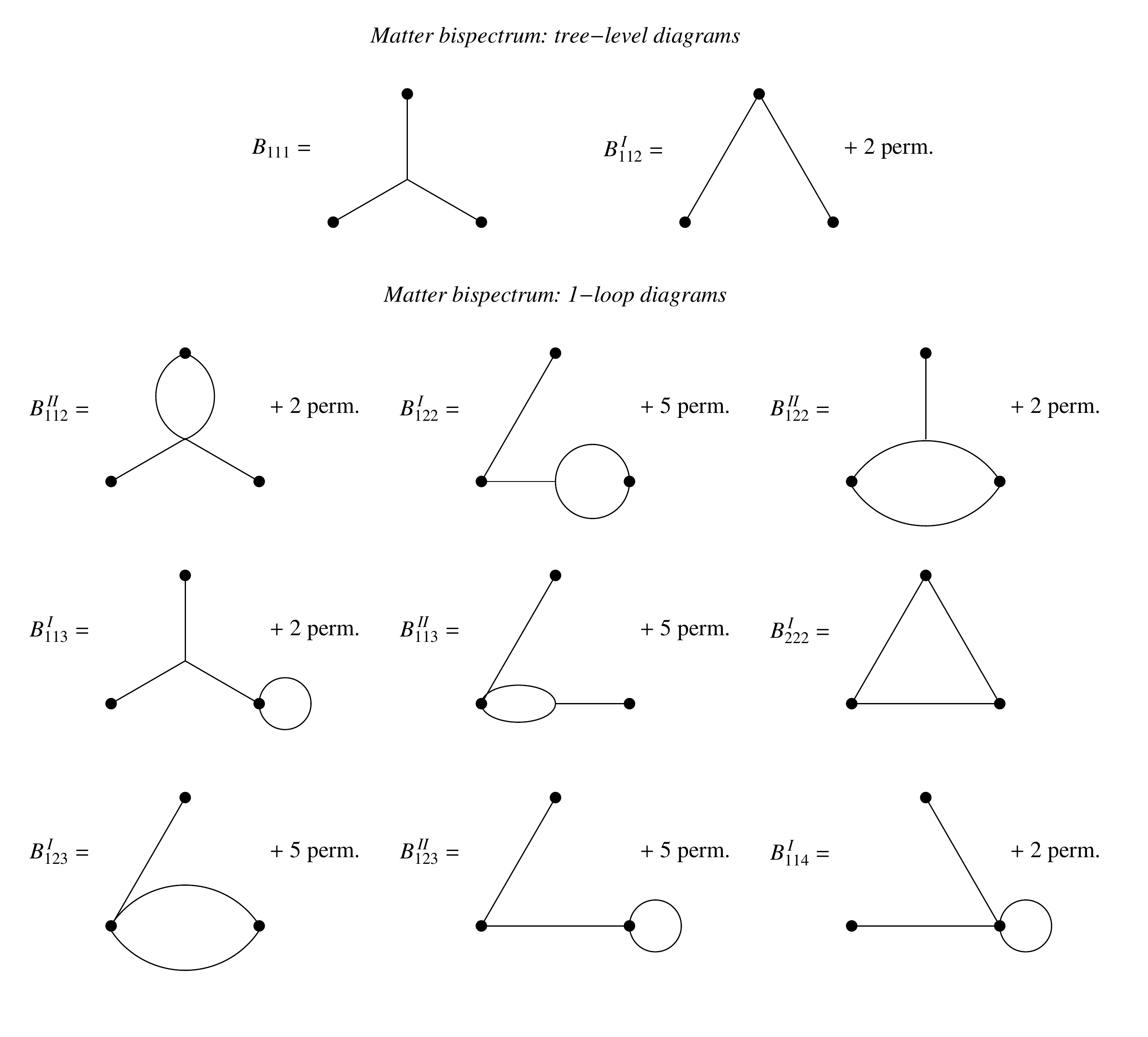}}
\caption{Diagrams representing the tree-level and 1-loop contributions to the matter bispectrum with generic non-Gaussian initial conditions.}
\label{fig:graphsBm}
\end{center}
\end{figure}

We consider now the non-linear matter bispectrum for non-Gaussian initial conditions. Similarly to power spectrum, we write the perturbative solution for the matter bispectrum up to corrections $\O(\d_0^7)$ as 
\beq
\label{eq:Bexp}
B(k_1,k_2,k_3) = B^{(3)}+B^{(4)}+B^{(5)}+B^{(6)}+\O(\d_0^7),
\eeq
where $B^{(p)}\sim\O(\d_0^p)$ and where all odd contributions vanish for Gaussian initial conditions.
Here $B^{(3)}=B_{111}\equiv B_0$ corresponds to the initial bispectrum, while higher-order terms are given by the following individual contributions
\bea
B^{(4)}(k_1,k_2,k_3)
& = &
B_{112}^I+B_{112}^{II},
\\
B^{(5)}(k_1,k_2,k_3)
& = &
B_{122}^I+B_{122}^{II}+B_{113}^I+B_{113}^{II},
\\
B^{(6)}(k_1,k_2,k_3)
& = &
B_{222}^I+B_{123}^{I}+B_{123}^{II}+B_{114}^{I}+{\rm 2\!\!-\!\!loop~terms},
\eea
where we make use of the notation for the individual terms on the r.h.s. of the equations above
\beq
\d_D(\kv_{123})B_{ijl}(k_1,k_2,k_3)\equiv\la\d_{\kv_1}^{(i)}\d_{\kv_2}^{(j)}\d_{\kv_3}^{(l)}\ra\,+\,{\rm cyc.},
\eeq
analogous to the one for the power spectrum. We can distinguish the tree-level expression
\beq\label{eq:B112I}
B_{112}^I  =  
2~F_2(\kv_1,\kv_2)~P_0(k_1)~P_0(k_2)+{\rm 2~perm.},
\eeq
from the 1-loop corrections
\bea
B_{112}^{II}
& = & 
\intq~ F_2(\qv,\kv_3-\qv)~T_0(\kv_1,\kv_2,\qv,\kv_3-\qv),
\label{eq:B112II}\\
B_{122}^{I}
&=&
2 ~P_0(k_1)\left[F_2(\kv_1,\kv_3)\intq~F_2(\qv,\kv_3-\qv)~B_0(k_3,q,|\kv_3-\qv|)+(k_3\leftrightarrow k_2)\right]+{\rm 2~perm.}
\nonumber\\
& = &
F_2(\kv_1,\kv_2)\left[P_0(k_1)~P_{12}(k_2)+P_0(k_2)~P_{12}(k_1)\right]
+{\rm 2~perm.},
\eea
\bea
B_{122}^{II}
&=&
4 \intq~F_2(\qv,\kv_2-\qv)~F_2(\kv_1+\qv,\kv_2-\qv)~B_0(k_1,q,|\kv_1+\qv|)~P_0(|\kv_2-\qv|) +{\rm 2~perm.},
\\
B_{113}^I
&=&
3B_0(k_1,k_2,k_3)\intq~F_3(\kv_3,\qv,-\qv)P_0(q)+{\rm 2~perm.},
\\
B_{113}^{II}
&=&
3 P_0(k_1)\intq~F_3(\kv_1,\qv,\kv_2-\qv)B_0(k_2,q,|\kv_2-\qv|)+(k_1\leftrightarrow k_2)+{\rm 2~perm.},
\\
B_{222}^I
&=&
8 \intq F_2(-\qv,\qv+\kv_1)F_2(-\qv-\kv_1,\qv-\kv_2)F_2(\kv_2-\qv,\qv)P_0(q)P_0(|\kv_1+\qv|)P_0(|\kv_2-\qv|),
\\
B_{123}^{I}
&=&
6~P_0(k_1) \intq ~F_3(\kv_1,\kv_2-\qv,\qv)~F_2(\kv_2-\qv,\qv)~P_0(|\kv_2-\qv|)~P_0(q)+{\rm 5~perm.},
\\
B_{123}^{II}
&=&
6~P_0(k_1)~P_0(k_2)~F_2(\kv_1,\kv_2)\int d^3q~ F_3(\kv_1, \qv,-\qv)~P_0(q)+{\rm 5~perm.}
\nonumber\\
&=&
F_2(\kv_1,\kv_2)\left[P_0(k_1)~P_{13}(k_2)+P_0(k_2)~P_{13}(k_1)\right]+{\rm 2~perm.},
\\
B_{114}^I
&=&
12~P_0(k_1)~P_0(k_2) \intq ~F_4(\qv,-\qv,-\kv_1,-\kv_2)~P_0(q)+{\rm 2~perm.}.
\eea 
Again, we ignored 2-loop corrections to the $B^{(6)}$ term. Here, the 1-loop contributions $B_{222}^I$, $B_{123}^I$, $B_{123}^{II}$ and $B_{114}^I$ are those derived by \citet{Scoccimarro1997} for Gaussian initial conditions, while the others, $B_{112}^{II}$, $B_{122}^I$, $B_{122}^{II}$, $B_{113}^{I}$ and $B_{113}^{II}$ depend on the initial bispectrum $B_0$ and trispectrum $T_0$. Initial correlation functions of order higher than the trispectrum appear only in 2-loop corrections and higher. In Fig.~\ref{fig:graphsBm} we show the diagrams representing the tree-level and 1-loop contributions to the matter bispectrum described by the expressions above. 

\begin{figure}[t]
\begin{center}
{\includegraphics[width=0.45\textwidth]{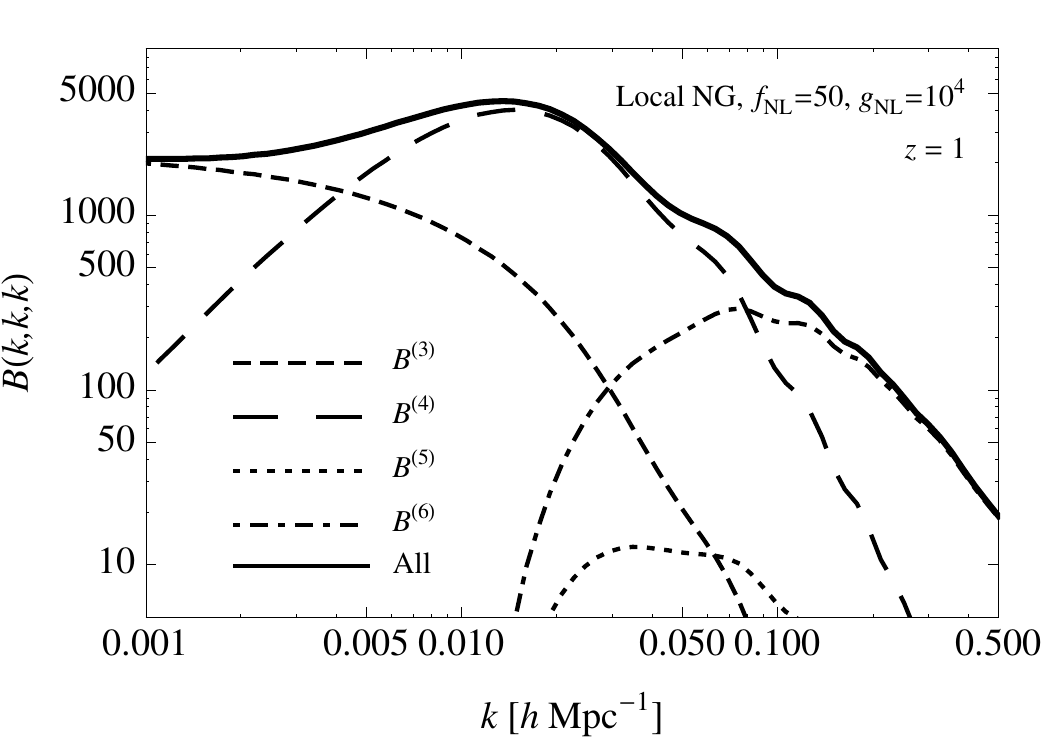}}
{\includegraphics[width=0.45\textwidth]{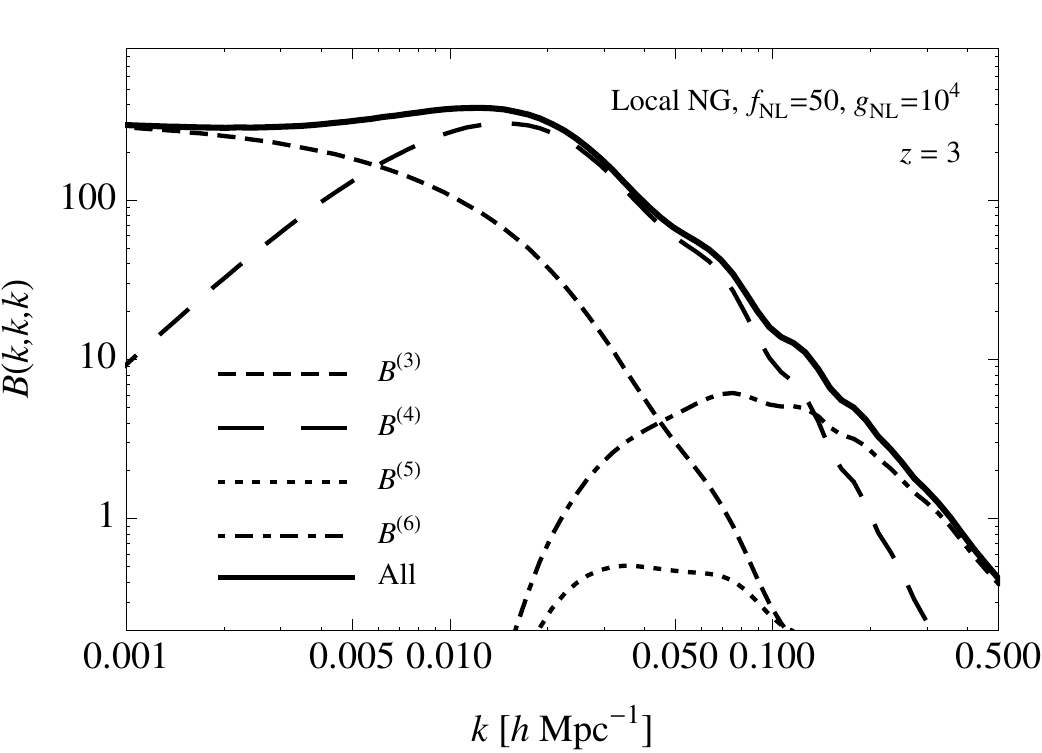}}
{\includegraphics[width=0.45\textwidth]{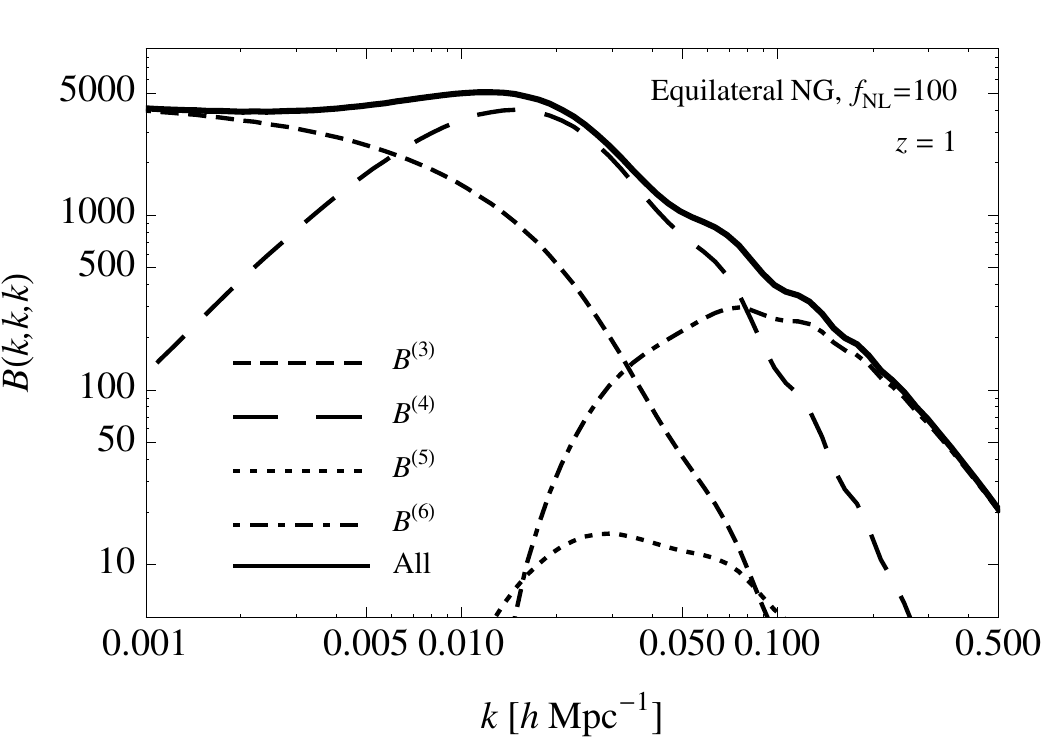}}
{\includegraphics[width=0.45\textwidth]{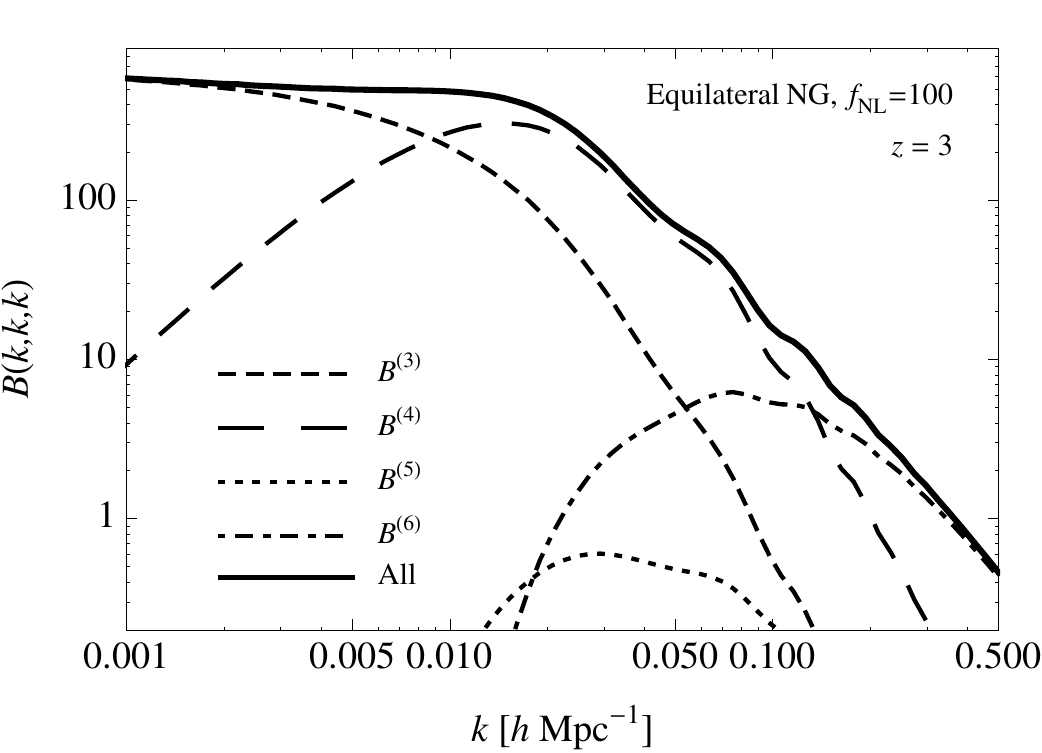}}
\caption{Equilateral configurations of the matter bispectrum, $B(k,k,k)$, for non-Gaussian initial conditions. {\it Upper panels:} local model with $\fNLl=50$ and $\gNL=10^4$ at redshifts $z=1$ ({\it left}) and $z=3$ ({\it right}). {\it Lower panels:} equilateral model with $\fNLl=100$ at redshifts $z=1$ ({\it left}) and $z=3$ ({\it right}). The main effect due to non-Gaussian initial conditions is represented by the primordial component $B^{(3)}$.}
\label{fig:bsmeq}
\end{center}
\end{figure}
In Fig.~\ref{fig:bsmeq} we show the equilateral configurations of the matter power spectrum $B(k,k,k)$ as a function of $k$ for non-Gaussian initial conditions. We choose values for the non-Gaussian parameters within current 1-$\sigma$ uncertainties. In the case of the local model we include the effect due to the initial trispectrum to 1-loop corrections, while for the equilateral model only a non-vanishing initial bispectrum is considered. The upper panels show the effect of local non-Gaussianity with $\fNLl=50$ and $\gNL=10^4$ at redshifts $z=1$ ({\it left}) and $z=3$ ({\it right}). The fourth-order term $B^{(4)}$ represented by long-dashed line, includes the correction due to the initial trispectrum $B_{112}^{II}$ of Eq.~(\ref{eq:B112II}) for the local case, which however, is completely negligible, so that the long-dashed line effectively indicates the tree-level term $B_{112}^I$ alone. The unnaturally large value for the parameter $\gNL$ is assumed only for illustrative purposes, since only for such large values we obtain a contribution comparable to the one from the term proportional to $\fNL^2$ in Eq.~(\ref{eq:Tlc}). In the lower panels of Fig.~\ref{fig:bsmeq} we consider the equilateral model with $\fNLe=100$ at redshifts $z=1$ ({\it left}) and $z=3$ ({\it right}). 

The $B^{(3)}\equiv B_0$ contribution, represented by short-dashed lines, is, by definition of the equilateral model itself, identical in the local and equilateral model for equilateral configurations, however the $B^{(5)}$ contribution is in principle different as it corresponds to integrals over a continuum of different configurations of the initial bispectrum. As it is the case for the power spectrum studied in \citep{TaruyaKoyamaMatsubara2008}, the extra 1-loop contributions corresponding to $B^{(5)}$ are quite suppressed with respect to the Gaussian component. Specifically, in the range of scales where the $B^{(5)}$ term is largest, roughly corresponding to $0.03$ to $0.05\kMpc$ in our cosmology, it represents a correction of a few percent to the Gaussian prediction, still smaller then the primordial contribution $B^{(3)}$ by nearly one order of magnitude for equilateral configurations. Notice that, for the equilateral model such configurations present the largest value for the primordial component, while this is not the case for the local model. Finally we notice that Gaussian 1-loop corrections in the $B^{(6)}$ term are significant, as expected, at somehow smaller scales, above $0.1\kMpc$ at redshift $z=1$ (see \citep{PanColesSzapudi2007,GuoJing2009} for recent simulations results with Gaussian initial conditions).

To study the dependence on the triangle shape, it is convenient to introduce a {\it reduced} matter bispectrum \citep{Fry1984}, defined as 
\beq
Q(k_1,k_2,k_3)\equiv\frac{B(k_1,k_2,k_3)}{\Sigma(k_1,k_2,k_3)},
\eeq
with
\beq
\Sigma(k_1,k_2,k_3)\equiv P(k_1)~P(k_2)+ {\rm 2~perm.}
\eeq
Its perturbative expansion in powers of $\d_0$ can be derived from Eq.~(\ref{eq:Bexp}) and from the series
\beq
\label{eq:Sexp}
\Sigma(k_1,k_2,k_3) = \Sigma^{(4)}+\Sigma^{(5)}+\Sigma^{(6)}+\O(\d_0^7),
\eeq
with
\bea
\Sigma^{(4)} & = & P^{(2)}(k_1)P^{(2)}(k_2)+{\rm 2~perm.},
\\
\Sigma^{(5)} & = & P^{(2)}(k_1)P^{(3)}(k_2)+{\rm 5~perm.},
\\
\Sigma^{(6)} & = & P^{(3)}(k_1)P^{(3)}(k_2)+{\rm 2~perm.}+P^{(2)}(k_1)P^{(4)}(k_2)+{\rm 5~perm.},
\eea
so that, with the same convention for the notation, we obtain
\beq
\label{eq:Qexp}
Q(k_1,k_2,k_3) = Q^{(-1)}+Q^{(0)}+Q^{(1)}+Q^{(2)}+\O(\d_0^3),
\eeq
where
\bea
Q^{(-1)} & = & \frac{B^{(3)}}{\Sigma^{(4)}},
\\
Q^{(0)} & = & \frac{B^{(4)}-Q^{(-1)}\Sigma^{(5)}}{\Sigma^{(4)}},
\\
Q^{(1)} & = & \frac{B^{(5)}-Q^{(0)}\Sigma^{(5)}-Q^{(-1)}\Sigma^{(6)}}{\Sigma^{(4)}},
\\
Q^{(2)} & = & \frac{B^{(6)}-Q^{(1)}\Sigma^{(5)}-Q^{(0)}\Sigma^{(6)}-Q^{(-1)}\Sigma^{(6)}}{\Sigma^{(4)}}.
\eea
The $Q^{(-1)}$ and $Q^{(1)}$ contributions vanish for Gaussian initial conditions. 

\begin{figure}[t]
\begin{center}
{\includegraphics[width=0.45\textwidth]{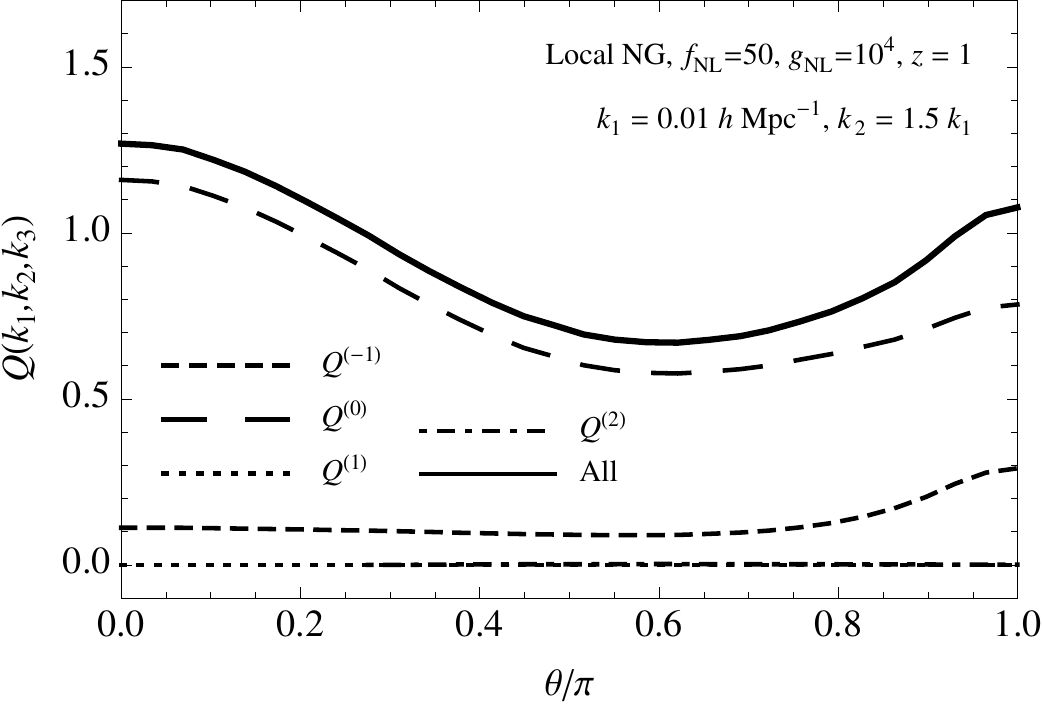}}
{\includegraphics[width=0.45\textwidth]{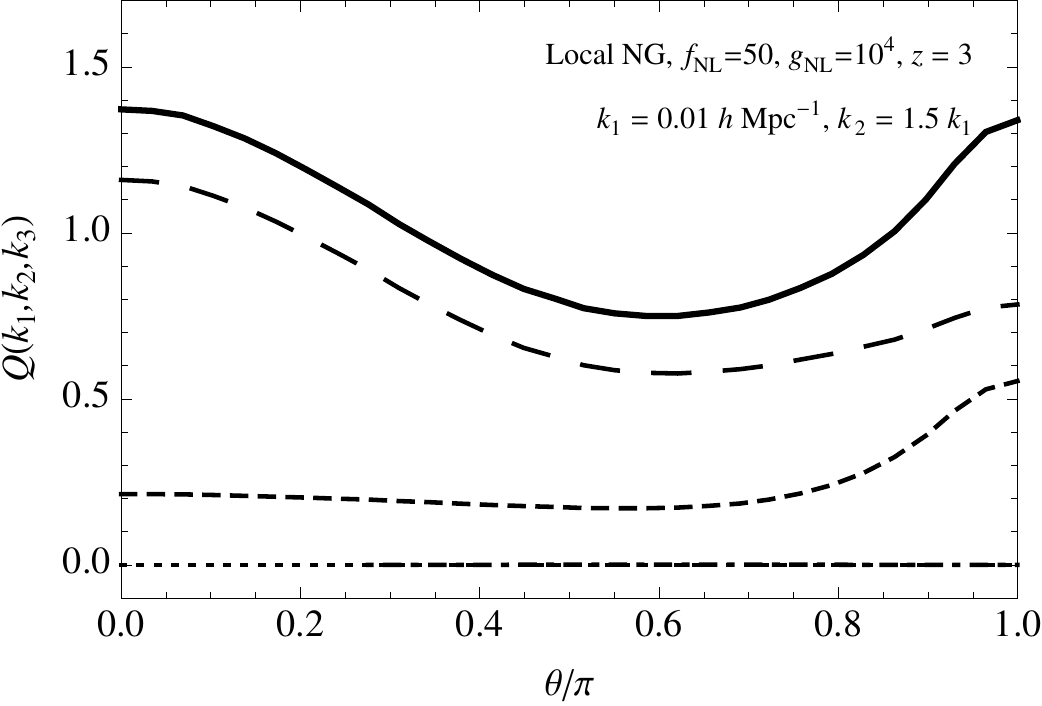}}
{\includegraphics[width=0.45\textwidth]{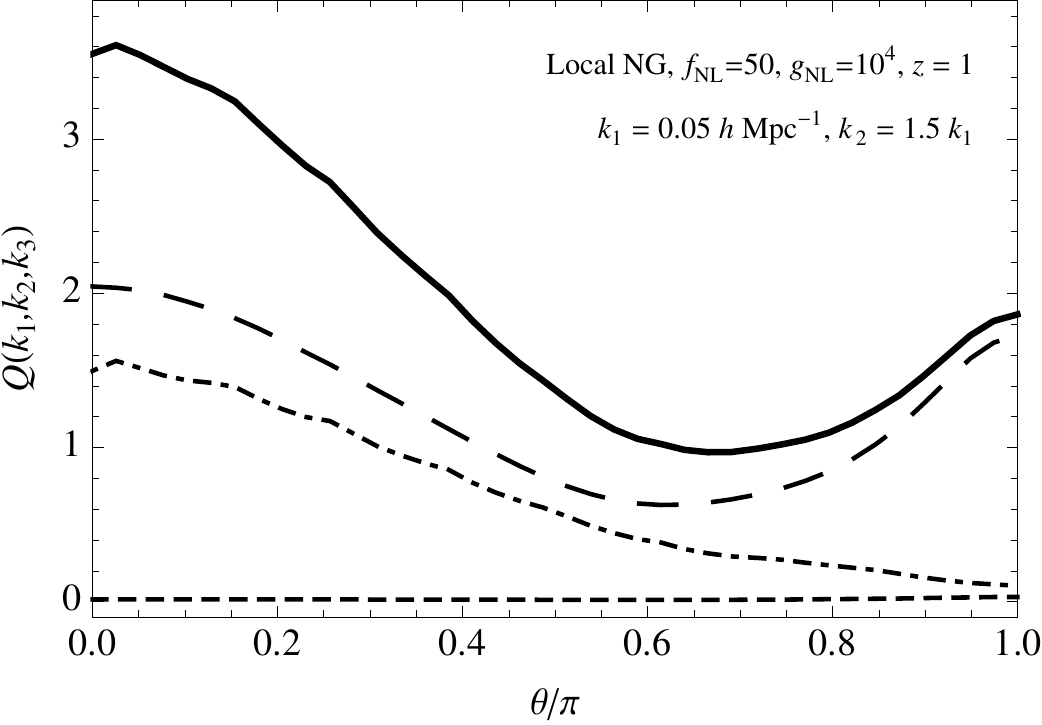}}
{\includegraphics[width=0.45\textwidth]{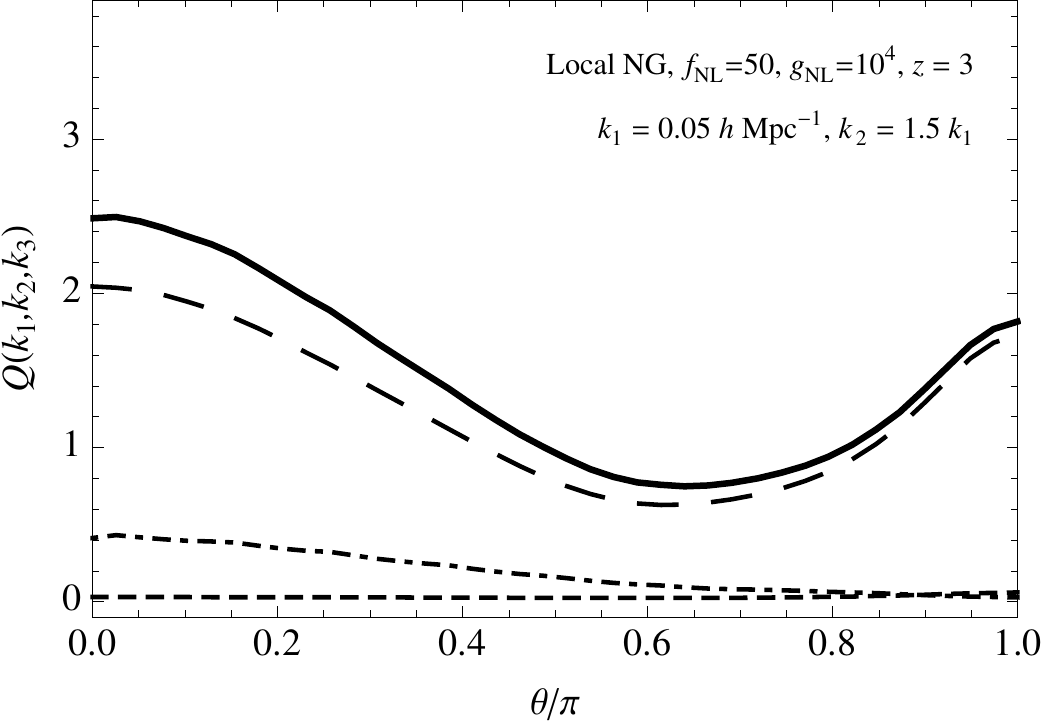}}
\caption{Reduced matter bispectrum $Q(k_1,k_2,k_3)$ with fixed $k_1$ and $k_2=1.5 k_1$ as a function of the angle $\theta$ between $\kv_1$ and $\kv_2$ for local non-Gaussian initial conditions with $\fNL=50$ and $\gNL=10^4$. Upper panels assume $k_1=0.01\kMpc$ while lower panels assume $k_1=0.05\kMpc$. Predictions in the left and right panels are evaluated respectively at redshift $z=1$ and $z=3$. As in Fig.~\ref{fig:bsmeq}, $Q^{(-1)}$ ({\it short-dashed line}), which includes the primordial component $B^{(3)}$, represents the main non-Gaussian contribution, which assumes the largest values for $\theta\simeq \pi$, corresponding to the {\it squeezed} limit for this set of configurations.}
\label{fig:bsmaLc}
\end{center}
\end{figure}

In Fig.~\ref{fig:bsmaLc} we show the reduced matter bispectrum $Q(k_1,k_2,k_3)$ with fixed $k_1$ and $k_2=1.5~ k_1$ as a function of the angle $\theta$ between $\kv_1$ and $\kv_2$ for non-Gaussian initial conditions of the local kind with $\fNL=50$ and non-vanishing initial bispectrum and trispectrum\footnote{Again, the contribution depending on the initial trispectrum $T_0$, even assuming $\gNL=10^4$, is completely negligible for all configurations shown in Fig.~\ref{fig:bsmaLc}.}. The upper panels assume $k_1=0.01\kMpc$, while lower panels assume $k_1=0.05\kMpc$. For the first set of configurations the leading correction to the $Q^{(0)}$ term ({\it long-dashed line}) is given by the primordial component  $Q^{(-1)}$ ({\it short-dashed line}), which assumes the largest values for $\theta\simeq \pi$, corresponding to the {\it squeezed} limit for this set of configurations. For the second set of configurations with $k_1=0.05\kMpc$ the leading correction to $Q^{(0)}$ is given instead by the 1-loop contributions in the term $Q^{(2)}$ ({\it dot-dashed line}), larger at small values of $\theta$ corresponding to large values of $k_3$. Predictions in the left and right panels are evaluated respectively at redshift $z=1$ and $z=3$. Choosing a value for the ratio $k_2/k_1$ closer to $1$ would lead to a much greater non-Gaussian component in the squeezed limit since, in fact, $Q^{(-1)}(k,k,k_3)\sim 1/k_3^2$ and therefore diverges for $\theta\rightarrow\pi$. We avoid in our examples such extreme cases of squeezed triangles since configurations with $k_1=k_2\gg k_3$ are difficult to measure in actual surveys due to the limited resolution in the wavenumber value.

\begin{figure}[t]
\begin{center}
{\includegraphics[width=0.45\textwidth]{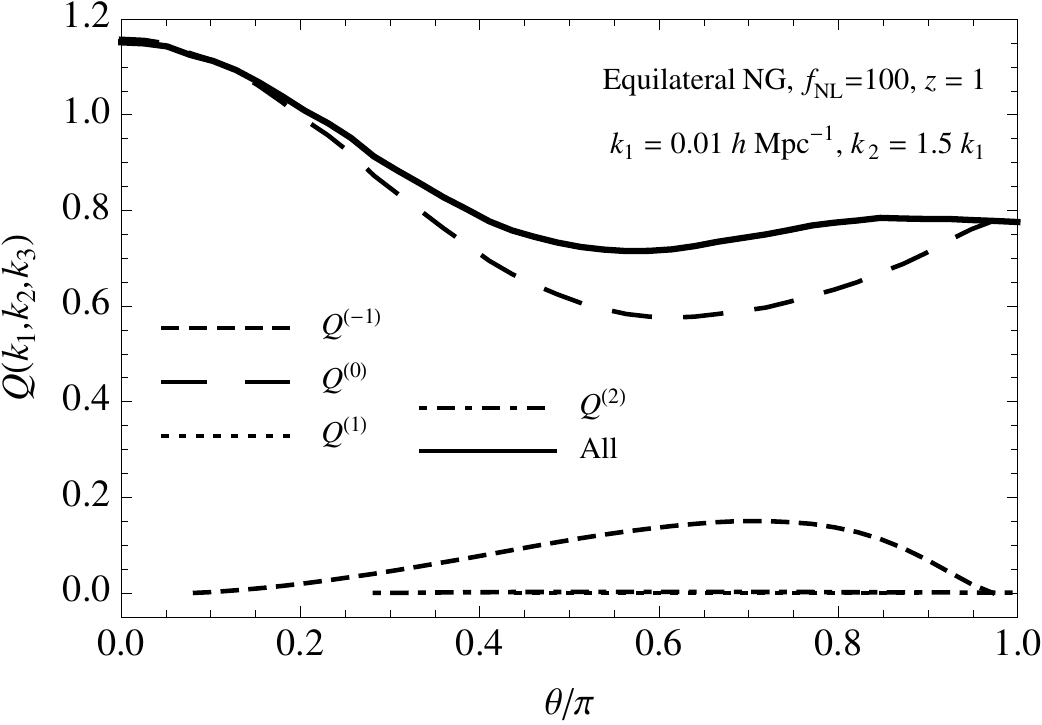}}
{\includegraphics[width=0.45\textwidth]{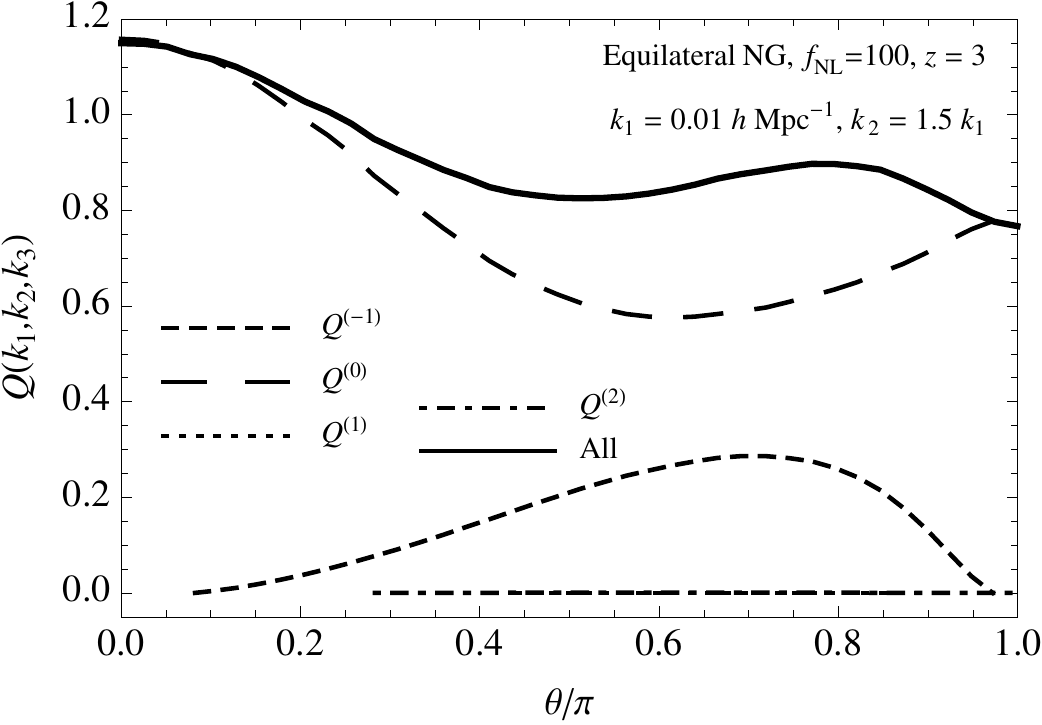}}
{\includegraphics[width=0.45\textwidth]{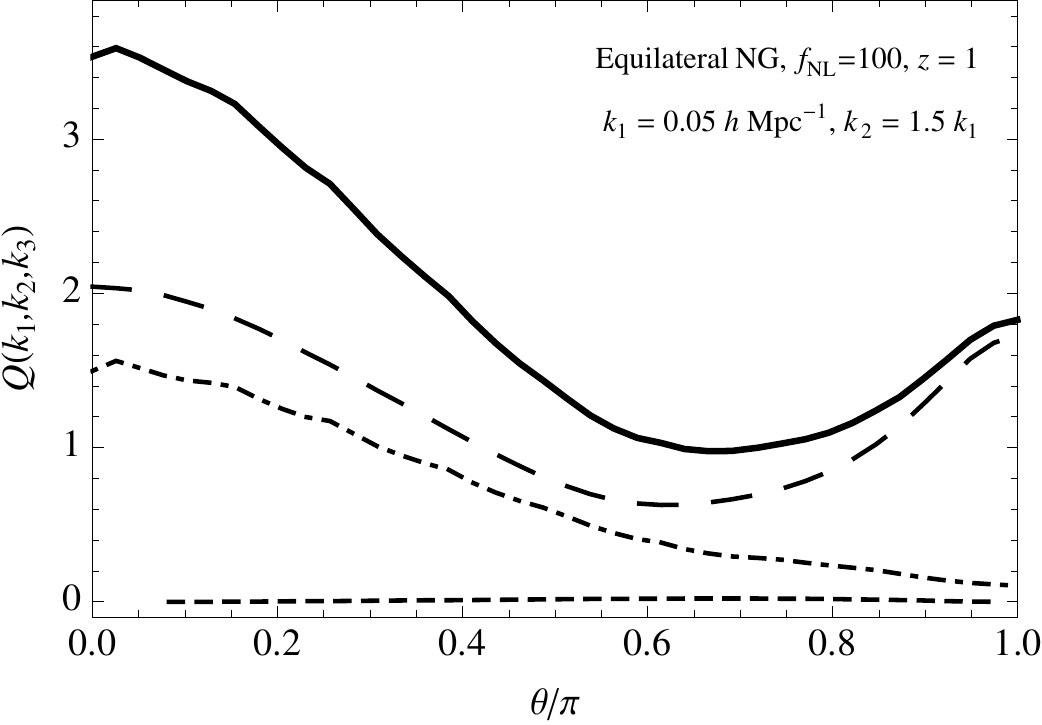}}
{\includegraphics[width=0.45\textwidth]{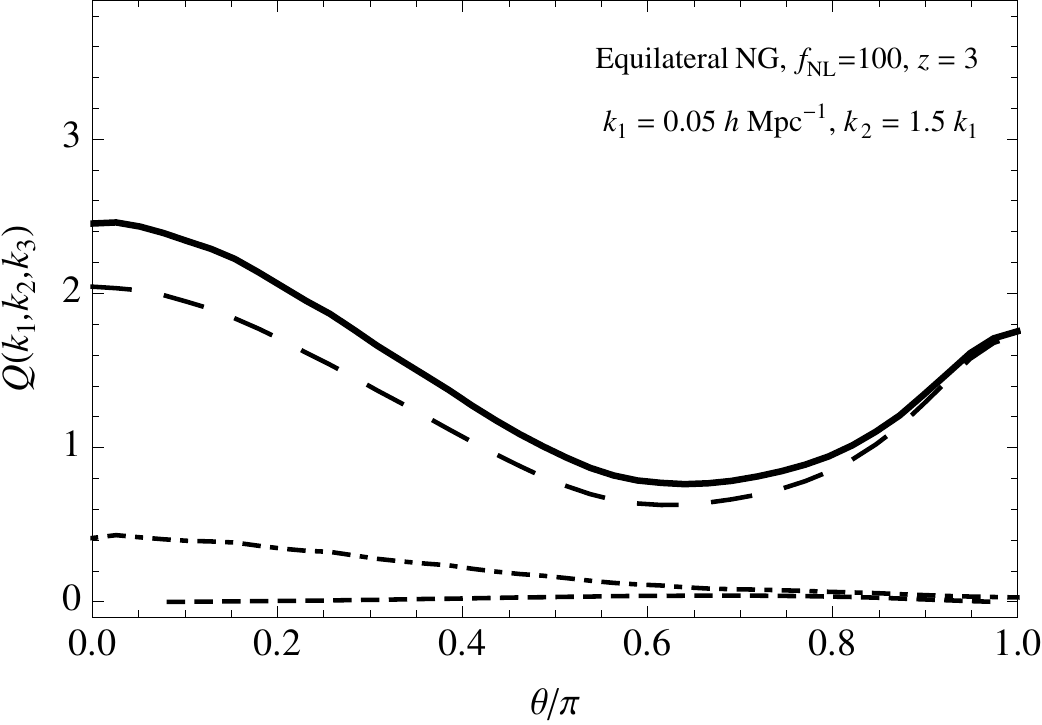}}
\caption{Same as Fig.~\ref{fig:bsmaLc} but for the equilateral non-Gaussian model with $\fNL=100$. In this case, values of $\theta\simeq 0.7\pi$ correspond to the equilateral triangle limit for this set of configurations.}
\label{fig:bsmaEq}
\end{center}
\end{figure}
Fig.~\ref{fig:bsmaEq} shows the same results as Fig.~\ref{fig:bsmaLc} but for the equilateral model with $\fNL=100$. In this case the largest non-Gaussian corrections correspond to the nearly equilateral configurations with $\theta\simeq 0.7\pi$. 

From these two examples it is particularly evident the different effect that alternative models of primordial non-Gaussianity can have on the matter bispectrum and that could lead, in the optimistic case of the detection of a large signal in future observation, to distinguish the specific shape-dependence of the initial component and to possibly identify the corresponding model.

\subsection{The matter trispectrum}
\label{ssec:trisp}

We consider here, finally, the expression for the matter trispectrum up to the same order in perturbation theory considered above for the bispectrum, that is $\O(\d_0^6)$, since it will be useful in the next section. We will focus our attention to the tree-level contributions, neglecting loop corrections. Following the notation introduced for the power spectrum, we can write
\beq
\label{eq:Texp}
T(\kv_1,\kv_2,\kv_3,\kv_4)= T^{(4)}+T^{(5)}+T^{(6)}+\O(\d_0^7),
\eeq
and, denoting individual terms as 
\beq
\d_D(\kv_{1234})T_{ijlm}(\kv_1,\kv_2,\kv_3,\kv_4)\equiv\la\d_{\kv_1}^{(i)}\d_{\kv_2}^{(j)}\d_{\kv_3}^{(l)}\d_{\kv_4}^{(m)}\ra\,+\,{\rm cyc.},
\eeq
we have, in the first place, the initial trispectrum $T^{(4)}=T_{1111}=T_0$, {\it i.e.} the connected four-point function, vanishing for Gaussian initial conditions, then
\beq
T^{(5)}(\kv_1,\kv_2,\kv_3,\kv_4)=T_{1112}^I+T_{1112}^{II}
\eeq
with $T_{1112}^I$ being the tree-level correction due to gravitational instability {\it and} a non-vanishing initial bispectrum $B_0$ given by 
\beq\label{eq:T1112}
T_{1112}^I(k_1,k_2,k_3,k_4) =2~B_0(k_1,k_2,k_{12})\left[F_2(\kv_{12},\kv_3)P_0(k_3)+F_2(\kv_{12},\kv_4)P_0(k_4)\right]+{\rm 5~perm.},
\eeq
while $T_{1112}^{II}$, which we do not write down explicitly, represents a 1-loop correction involving the connected five-point function, and finally
\beq
T^{(6)}(\kv_1,\kv_2,\kv_3,\kv_4)=T_{1122}^I+T_{1113}^{I}+{\rm 1\!\!-\!\!loop~and~2\!\!-\!\!loop~terms},
\eeq
where 
\bea
T_{1122}^I(k_1,k_2,k_3,k_4) 
& = & 
4~F_2(\kv_{13},-\kv_1)~F_2(\kv_{13},\kv_2)P_0(k_1)P_0(k_2)P_0(k_{13})+{\rm 11~perm.},
\\
T_{1113}^I(k_1,k_2,k_3,k_4) 
& = & 
6~F_3(\kv_{1},\kv_2,\kv_3)P_0(k_1)P_0(k_2)P_0(k_{3})+{\rm 3~perm.},
\eea
represent the leading tree-level contribution for Gaussian initial conditions. Again we neglect all 1-loop and 2-loop corrections depending, in general, on the square of $B_0$, on products of $P_0$ and $T_0$ and on higher-order initial correlation functions. 

\begin{figure}[t]
\begin{center}
{\includegraphics[width=1\textwidth]{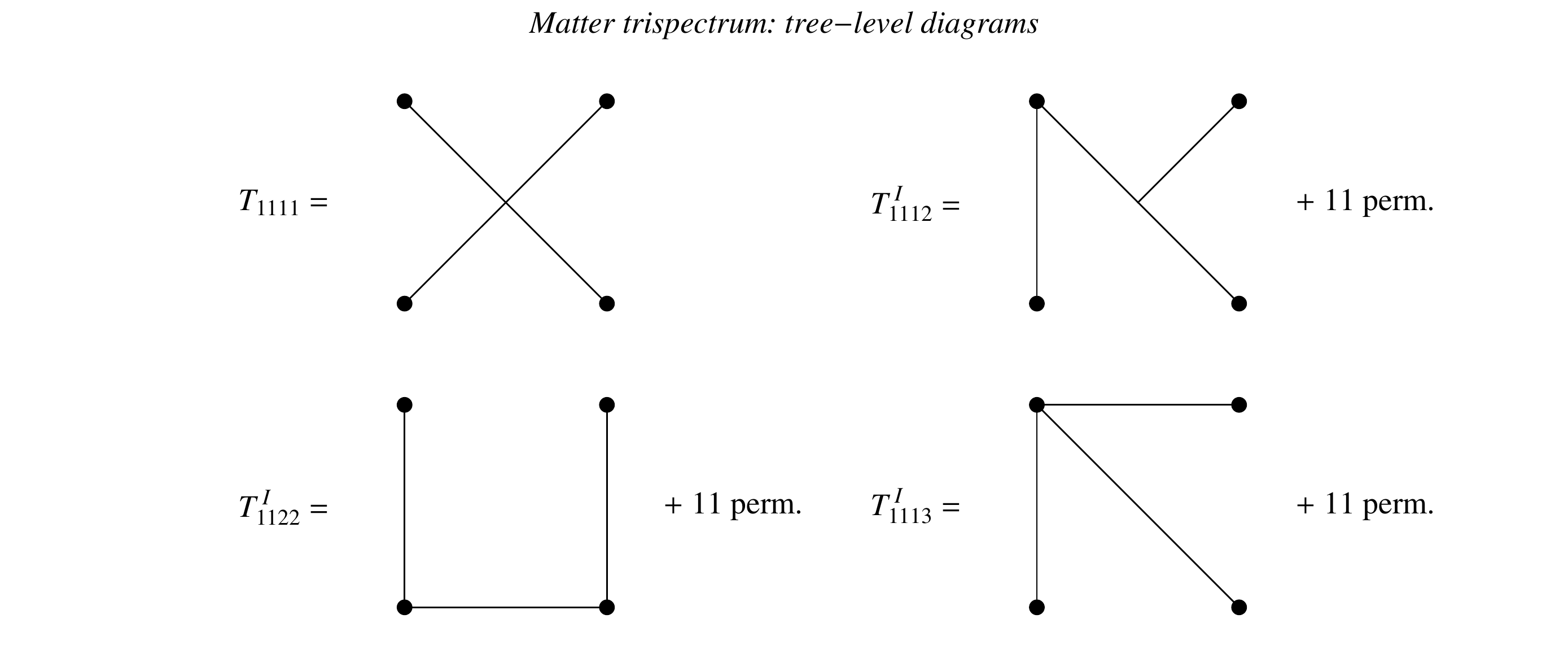}}
\caption{Diagrams representing the tree-level contributions to the matter trispectrum with generic non-Gaussian initial conditions.}
\label{fig:graphsTm}
\end{center}
\end{figure}

In Fig.~\ref{fig:graphsTm} we show the diagrams representing the tree-level contributions $T_{1111}$, $T_{1112}$, $T_{1122}^I$ and $T_{1113}^I$ to the matter trispectrum as described above. 

We will show in the next section that the primordial component $T_0=T_{1111}$ and the $T_{1112}^I$ term, arising from a combination of primordial non-Gaussianity and non-linearities in the matter evolution, will contribute significantly to the 1-loop corrections to the galaxy bispectrum. As we do not consider, for the time being in this paper, all 2-loop corrections we ignore consistently 1-loop corrections to the matter trispectrum as they will be part of contributions to the galaxy bispectrum involving one further integration, therefore effectively corresponding to 2-loop corrections in terms of the linear density field.

\section{Galaxy correlators in the local bias model and non-Gaussian initial conditions}
\label{sec:galaxyPT}

\citet{TaruyaKoyamaMatsubara2008} derived an perturbative expression for the galaxy power spectrum in presence of non-Gaussian initial conditions following the local bias prescription of \citet{FryGaztanaga1993}. They have shown that an effect on the linear halo bias similar to the one measured in N-body simulations \citep{DalalEtal2008,DesjacquesSeljakIliev2009,PillepichPorcianiHahn2008,GrossiEtal2009} in its redshift and scale dependence can be derived as a 1-loop correction to the galaxy power spectrum which is of order $\O(\d_0^3)$ in the linear density field $\d_0$. Here we follow the same approach, based on the assumption that relation between the smoothed galaxy overdensity $\d_g(\xv)$ and the smoothed density contrast $\d_{R}(\xv)$ is {\it local}, that is
\beq
\d_{g,R}(\xv)=f[\d_R(\xv)].
\eeq
Assuming small fluctuations at large scales one can Taylor-expand the relation above as
\beq\label{eq:LocalBias}
\d_{g,R}(\xv)=b_1\d_R(\xv)+\frac{b_2}{2}[\d_R^2(\xv)-\la\d_R^2(\xv)\ra]+\frac{b_3}{6}[\d_R^3(\xv)-\la\d_R^3(\xv)]+...,
\eeq
where $b_i$ are constant bias parameters. In Fourier space, for $k\ne 0$, we can write the filtered galaxy overdensity as
\beq
\d_{g,R}(\kv)=\sum_{n=1}^{\infty}\frac{b_n}{n!}\d_{g,R}^{(n)}(\kv),
\eeq
where
\beq
\d_{g,R}^{(n)}(\kv)=\int\!\!d^3q_1\cdots d^3q_n\d_D(\kv\!-\!\qv_{1...n})\d_R(\qv_1)\cdots\d_R(\qv_n),
\eeq
with $\d_R(\kv)\equiv W_R(k)\d_{\kv}$, $W_R$ being an appropriate smoothing function. From a comparison between these expressions and Eq.s~(\ref{eq:PT}) and (\ref{eq:PTterm}), it is evident that perturbative corrections to the correlators of the galaxy distributions can be obtained in a similar way as those for matter correlators considered in the previous section.

In this section, we compute the galaxy bispectrum up order $\O(\d_0^5)$ in perturbation theory, excluding therefore 1-loop corrections to the bispectrum for Gaussian initial conditions---which are of order $\O(\d_0^6)$---but including 1-loop corrections involving the initial bispectrum, and, for the local non-Gaussian model, the initial trispectrum. This amounts to neglect non-linear effects at small scales while concentrating our attention on the large-scale contributions. For the filter $W_R$ we assume a top-hat window function in position space with $R=5\Mpc$, although this value, used for the quantities shown in the figures, is not strictly relevant to the main results of 
the paper.

\subsection{The galaxy power spectrum}
\label{sec:GalaxyPS} 
 
We begin by summarizing the results of \citet{TaruyaKoyamaMatsubara2008} on the galaxy power spectrum in the framework of the local bias model of Eq.~(\ref{eq:LocalBias}) and non-Gaussian initial conditions that will be useful for our analysis of the galaxy bispectrum. Following \citet{SmithScoccimarroSheth2007} and \citet{TaruyaKoyamaMatsubara2008}, we define the unfiltered galaxy power spectrum $P_g(k)$ in terms of the power spectrum of the smoothed density field $P_{g,R}(k)$ simply as $P_g(k)\equiv{P_{g,R}(k)}/{W^2(kR)}$. The perturbative expression for the galaxy power spectrum, consistent with the expansion of Eq.~(\ref{eq:LocalBias}), can be written as
\beq
P_g(k)\equiv\frac{P_{g,R}(k)}{W^2(kR)}=P_{g,11}+P_{g,12}+P_{g,22}+P_{g,13}+\dots,
\eeq 
where each term on the r.h.s. is defined according to the notation 
\beq\label{eq:gPTterms}
\d_D(\kv_{12})P_{g,ij}(k_1)
\equiv
\frac{b_i}{i!}\frac{b_j}{j!}\frac{\la\d_{R}^{i}(\kv_1)\d_{R}^{j}(\kv_2)\ra\,+\,\la\d_{R}^{j}(\kv_1)\d_{R}^{i}(\kv_2)\ra}{W(k_1R)W(k_2R)}.
\eeq
We choose therefore to denote each contribution to the galaxy power spectrum, and later to the galaxy bispectrum, in terms of their dependence on the bias parameters $b_i$, instead of their perturbative order, keeping in mind, however, that, in general, $P_{g,ij}$ contains terms $\O(\d_0^{i+j})$ and higher. 
The first contributions, involving corrections up to $\O(\d_0^5)$, can be written in terms of the {\it non-linear} matter correlators $P$, $B$ and $T$ as \citep{McDonald2006,SmithScoccimarroSheth2007,TaruyaKoyamaMatsubara2008},
\bea
P_{g,11}
& = & b_1^2P(k),
\\
P_{g,12}
& = & 
b_1b_2\intq~\tW(\qv,\kv\!-\!\qv)~B(k,q,|\kv\!-\!\qv|),
\\
P_{g,22}^I
& = & 
\frac{b_2^2}{2}\intq~\tW^2(\qv,\kv\!-\!\qv)~P(q)~P(|\kv\!-\!\qv|),
\\
P_{g,22}^{II}
& = & 
\frac{b_2^2}{4}\intq d^3p~\tW(\qv,\kv\!-\!\qv)~\tW(\pv,-\kv\!-\!\pv)~T(\qv,\kv\!-\!\qv,\pv,-\kv\!-\!\pv),
\\
P_{g,13}^I
& = & 
b_1b_3~ P(k)\intq~W^2(\qv)~P(q),
\\
P_{g,13}^{II}
& = & 
\frac{b_1b_3}{3}\intq d^3p~\tW(\qv,\pv)~\tW(\qv\!+\!\pv,-\kv\!-\!\qv\!-\!\pv)~T(\kv,\qv,\pv,-\kv\!-\!\qv\!-\!\pv),
\eea
where we introduced, for convenience, the combination of window functions
\beq
\tilde{W}_R(\qv_1,\qv_2)\equiv\frac{W(q_1R)W(q_2R)}{W(q_{12}R)}.
\eeq
Clearly, the quantities $P_{g,22}$ and $P_{g,13}$ as defined in Eq.~(\ref{eq:gPTterms}) correspond to the sums $P_{g,22}^{I}+P_{g,22}^{II}$ and $P_{g,13}^{I}+P_{g,13}^{II}$, respectively. It is as well evident that the diagrams in Fig.~\ref{fig:graphsPm} can represent as well the terms in the perturbative expansion of the galaxy power spectrum once we replace the mode couplings in each node with the powers of the series in Eq.~(\ref{eq:LocalBias}) and the primordial correlators with those of the non-linear matter density field.

We can rewrite the expressions above making explicit the perturbative contributions to the matter power spectrum, bispectrum and trispectrum to keep track of the order of the expansion in terms of the initial density field $\d_0$. Excluding  corrections of order $\O(\d_0^5)$ we have
\bea
P_{g,11}
& \simeq & b_1^2\left[P^{(2)}+P^{(3)}+P^{(4)}\right],
\\
P_{g,12}
& \simeq & 
b_1b_2\intq~\tW(\qv,\kv\!-\!\qv)~[B^{(3)}(k,q,|\kv\!-\!\qv|)+B^{(3)}(k,q,|\kv\!-\!\qv|)],
\label{eq:Pg12}
\\
P_{g,22}^I
& \simeq & 
\frac{b_2^2}{2}\intq~\tW^2(\qv,\kv\!-\!\qv)~P^{(2)}(q)~P^{(2)}(|\kv\!-\!\qv|),
\\
P_{g,22}^{II}
& \simeq & 
\frac{b_2^2}{4}\intq d^3p~\tW(\qv,\kv\!-\!\qv)~\tW(\pv,-\kv\!-\!\pv)~T^{(4)}(\qv,\kv\!-\!\qv,\pv,-\kv\!-\!\pv),
\\
P_{g,13}^I
& \simeq & 
b_1b_3 ~P^{(2)}(k)\intq~W^2(\qv)~P^{(2)}(q),
\\
P_{g,13}^{II}
& \simeq & 
\frac{b_1b_3}{3}\intq d^3p~\tW(\qv,\pv)~\tW(\qv\!+\!\pv,-\kv\!-\!\qv\!-\!\pv)~T^{(4)}(\kv,\qv,\pv,-\kv\!-\!\qv\!-\!\pv).
\eea
For consistency with the previous section and for simplicity we will ignore all 2-loop contributions. The interesting term in this expansion is given by the $\O(\d_0^3)$ correction, 
\beq
P_{g,12}^{(3)}(k)
 =  
b_1b_2\intq~\tW(\qv,\kv\!-\!\qv)~B_0(k,q,|\kv\!-\!\qv|),
\eeq
which depends on the initial matter bispectrum $B^{(3)}=B_0$. In fact, assuming the expression for the initial bispectrum of the local model given by Eq.~(\ref{eq:Blc}), in the large-scale limit $k\rightarrow 0$ we have
\bea
P_{g,12}^{(3)}(k)
& = &
2\fNL b_1b_2\intq~\tW(\qv,\kv\!-\!\qv)~M(k,z)M(q,z)M(|\kv\!-\!\qv|,z)
\nonumber\\
& & \times\left[P_\Phi(k)P_\Phi(q)+P_\Phi(k)P_\Phi(|\kv\!-\!\qv|)+P_\Phi(q)P_\Phi(|\kv\!-\!\qv|)\right]
\nonumber\\
& \stackrel{k\rightarrow 0}{\simeq} &
4\fNL b_1b_2 M(k,z)P_\Phi(k)\intq~W_R^2(\qv)~M^2(q,z)
P_\Phi(q)
\nonumber\\
& = &
4\fNL b_1b_2 M(k,z)P_\Phi(k)\sigma_R^2(z),
\eea
which can be written as
\beq
P_{g,12}^{(3)}(k) = 
4\fNLl \frac{b_2}{b_1} \sigma_R^2(z)\frac{1}{M(k,z)}P_{g,11}^{(2)}(k),
\eeq
corresponding to Eq.~(5.3) in \citep{TaruyaKoyamaMatsubara2008}. Excluding corrections of the order $\O(\d^4)$, the galaxy bispectrum is given by
\beq
P_g(k)\simeq P_{g,11}^{(2)}+P_{g,11}^{(3)}+P_{g,12}^{(3)},
\eeq
and, neglecting the small term $P_{g,11}^{(3)}=b_1^2 P_{12}$, we can write
\beq\label{eq:PgTKM}
P_g(k)\simeq b_1^2\left[1+4\fNLl \frac{b_2}{b_1} \sigma_R^2(z)\frac{1}{M(k,z)}  \right]P_0(k),
\eeq
and the $P_{g,12}$ contribution can be expressed as a scale-dependent, non-Gaussian correction to the linear bias, defined as 
\beq
b_1(\fNL)=b_1(\fNL=0)+\Delta b_1(\fNL,k),
\eeq
and given by
\beq\label{eq:dbTKM}
\Delta b_1=2\fNLl~b_2~\sigma_R^2(z) \frac{1}{M(k,z)}.
\eeq
As pointed-out by \citet{TaruyaKoyamaMatsubara2008}, this expression is similar in its scale and redshift dependence to the correction derived in the peak-background split framework by \citep{DalalEtal2008,SlosarEtal2008,AfshordiTolley2008}, from a model for high-peak correlators by \citep{MatarreseVerde2008} and in an even different derivation, but related to the local bias assumption, in \citep{McDonald2008}. In particular, the peak-background split approach, \citep{DalalEtal2008,SlosarEtal2008,AfshordiTolley2008} provides, for the correction to the linear bias of {\it halos} of mass $M$, $b_{h,1}(M,z)$, the expression
\beq\label{eq:dblocal}
\Delta b_{h,1}= 2 \fNLl[b_{h,1}(M,z)-1]\delta_c\frac{1}{M(k,z)},
\eeq
where $\d_c=1.686$ is the critical threshold for spherical collapse. 

Following a different approach, \citet{MatarreseVerde2008} considered the peak number density defined by the expression $\rho_{p,M}(\xv)=\theta[\d_R(\xv)-\delta_c]$, where $\theta$ is a step function. Based on earlier works by \citet{GrinsteinWise1986} and \citet{MatarreseLucchinBonometto1986} which provide an expression for the peak two-point function of the form
\beq
\xi_p(\xv_1,\xv_2)=\exp[X(\xv_1,\xv_2)]-1,
\eeq
where the $X(\xv_1,\xv_2)$ represents a series involving all higher-order correlators of the initial density field, they derive a correction to the peak correlation function in terms of the three-point correlation function. Their result can be rewritten in the form of a {\it peak} linear bias correction given by, in our notation
\beq\label{eq:dbMV}
\Delta b_{p,1} = \frac{\nu^2}{\sigma_R^2}\frac{1}{2P_0(k)}\int d^3q~ \tW(\qv,\kv-\qv) B_0(k,q,|\kv-\qv|)
\eeq
where $\nu\equiv\d_c/\sigma_R$. This expression is valid in the high-peak limit $\nu/\sigma_R\gg 1$ but, as the one derived by \citet{TaruyaKoyamaMatsubara2008}, is quite general since it applies to any form of the initial bispectrum $B_0$. In the large-scale limit and for local non-Gaussianity we can find the asymptotic behavior for the, Lagrangian, peak bias correction 
\beq
\Delta b_{p,1} \stackrel{k\rightarrow 0}{\simeq} 2\fNL\frac{\nu^2}{\sigma_R^2}\frac{\sigma_R^2}{M(k)}=2\fNL(b_{p,1}-1)\delta_c\frac{1}{M(k)},
\eeq
where the last equality shows the equivalence with Eq.~(\ref{eq:dblocal}) after identifying $b_{p,1}=1+\nu/\sigma_R$ as the Eulerian linear peak bias.

It is interesting to notice that, if we interpret the bias parameters introduced by the local bias model of Eq.~(\ref{eq:LocalBias}) as {\it halo bias functions}, $b_{h,i}(M)$, and if we assume for these quantities the {\it Gaussian} predictions of the Halo Model, \citep{MoWhite1996,ScoccimarroEtal2001A} then, in the high-threshold limit $\nu/\sigma_R\gg 1$, we can approximate
\beq
b_2~\sigma_R^2\simeq [b_{h,1}-1]~\delta_c\simeq \nu^2=\frac{\delta_c^2}{\sigma_R^2},
\eeq
so that all these different bias corrections, including Eq.~(\ref{eq:dbTKM}), provide similar quantitative results. In fact, the factor $\nu^2/\sigma_R^2$ in Eq.~(\ref{eq:dbMV}), could just as well be interpreted as a non-linear, quadratic bias parameter. 

\citet{TaruyaKoyamaMatsubara2008} notice how the presence of the $\sigma_R^2$ term in the asymptotic expression of Eq.~(\ref{eq:dbTKM}) might represent a problem due to its logarithmic diverge as $R\rightarrow 0$ in the case of local non-Gaussianity. However, if we are to assume that indeed such different results describe the same effect, the peak-background split approach would provide a well defined prediction for the product $b_2\sigma_R^2$ when $b_2$ describes quadratic non-linearities of the halo bias relation. On the other hand, Eq.~(\ref{eq:dblocal}) has been shown to well describe numerical results, \citep{DesjacquesSeljakIliev2009,PillepichPorcianiHahn2008,GrossiEtal2009}, up to a corrective factor $q=0.75$, possibly related to ellipsoidal collapse \citep{GrossiEtal2009}. They also notice that other difficulties arise for negative values of $\fNLl$ or $b_2$, since in those cases, in the large-scale limit, the galaxy power spectrum in Eq.~(\ref{eq:PgTKM}) becomes negative. Such problem, however might just be an artifact of the truncated perturbative expansion rather than a pathology of the local model of non-Gaussianity. 

We will not discuss further all peculiar approximations involved in the different approaches proposed in the literature. As we will se in the next Section, large-scale behaviors proportional the quantity $\sigma_R$ are present
for several corrections to the galaxy bispectrum. We will therefore avoid interpreting our results as {\it quantitative} predictions limiting ourselves to present them as a derivation of the functional forms that we can expect for non-Gaussian contributions to the galaxy bispectrum.

\subsection{The galaxy bispectrum}

We study now the perturbative corrections to the galaxy bispectrum under the assumption of local bias, which represent the main results of this paper. Similarly to the case of the power spectrum, we can define the unfiltered galaxy bispectrum as
\beq
B_{g}(k_1,k_2,k_3)
\equiv 
\frac{B_{g,R}(k_1,k_2,k_3)}{W(k_1R)W(k_2R)W(k_3R)},
\eeq
which we can expand as
\beq
\label{eq:BgExp}
B_{g}=
B_{g,111}+B_{g,112}+B_{g,122}+B_{g,113} +\ldots,
\eeq
where each term can be formally defined as
\beq\label{eq:gPTtermsB}
\d_D(\kv_{123})B_{g,ijl}(k_1,k_2,k_3)
\equiv
\frac{b_i}{i!}\frac{b_j}{j!}\frac{b_l}{l!}\frac{\la\d_{R}^{i}(\kv_1)\d_{R}^{j}(\kv_2)\d_{R}^{l}(\kv_3)\ra\,+\,{\rm perm.}}{W(k_1R)W(k_2R)W(k_3R)}.
\eeq
The contributions containing terms up to $\O(\d_0^6)$, as a function of the {\it non-linear} matter correlators $P$, $B$ and $T$ are
\bea
B_{g,111}
& = &   
b_1^3 B,
\label{eq:Bg111}
\\
B_{g,112}^I
& = &   
b_1^2b_2~\tW(\kv_1,\kv_2)~P(k_1)~P(k_2)+{\rm 2~perm.},
\label{eq:Bg112I}
\\
B_{g,112}^{II}
& = &   
\frac{b_1^2b_2}{2}\intq~\tW(\qv,\kv_3\!-\!\qv)~T(\kv_1,\kv_2,\qv,\kv_3\!-\!\qv)+{\rm 2~perm.},
\label{eq:Bg112II}
\\
B_{g,122}^I
& = &
\frac{b_1b_2^2}{2} 
~\tW(\kv_1,\kv_3)~P(k_1)\intq~\tW(\qv,\kv_3\!-\!\qv)~B(k_3,q,|\kv_3\!-\!\qv|)+{\rm 5~perm.},
\label{eq:Bg122I}
\\
B_{g,122}^{II}
& = &
b_1b_2^2\intq~\tW(\qv,\kv_2\!-\!\qv)~\tW(\kv_1\!+\!\qv,\kv_2\!-\!\qv)~B(k_1,q,|\kv_1\!+\!\qv|)~P(|\kv_2\!-\!\qv|)+{\rm 2~perm.},
\label{eq:Bg122II}
\\
B_{g,113}^I
& = &   
\frac{b_1^2b_3}{2}~\tW(\kv_1,\kv_2)~P(k_1)\intq~\tW(\qv,\kv_2\!-\!\qv)~B(k_2,q,|\kv_2\!-\!\qv|)+{\rm 5~perm.},
\label{eq:Bg113I}
\\
B_{g,113}^{II}
& = &   
\frac{3}{2}~b_1^2b_3~ B(k_1,k_2,k_3)\intq~W_R^2(q)~P(q),
\label{eq:Bg113II}
\\
B_{g,222}^I
&=&
\frac{b_2^3}{2} \intq \tW(-\qv,\qv+\kv_1)\tW(-\qv-\kv_1,\qv-\kv_2)\tW(\kv_2-\qv,\qv)P(q)P(|\kv_1+\qv|)P(|\kv_2-\qv|),
\\
B_{g,123}^{I}
&=&
\frac{b_1b_2b_3}{2}\tW(\kv_1\kv_2)P(k_1) \intq \tW^2(\qv,\kv_2-\qv)P(|\kv_2-\qv|)P(q)+{\rm 5~perm.},
\\
B_{g,123}^{II}
&=&
b_1b_2b_3 \tW(\kv_1,\kv_2)P(k_1)P(k_2)\int d^3q W_R^2(q)P(q)+{\rm 2~perm.},
\label{eq:Bg123II}
\\
B_{g,114}^I
&=&
\frac{b_1^2b_4}{2}\tW(\kv_1,\kv_2)P(k_1)P(k_2) \intq W^2_R(q)P(q)+{\rm 2~perm.},
\label{eq:Bg114I}
\eea
where we ignored 2-loop contributions in the terms $B_{g,122}$ and $B_{g,113}$ which depend on the matter five-point function and 2- and 3-loop contributions in the terms $B_{g,222}$, $B_{g,123}$ and $B_{g,114}$ depending on squares of $B$, products of $P$ and $T$ and on the matter six-point function. 

The 1-loop corrections in $B_{g,222}^{I}$, $B_{g,123}^{I}$, $B_{g,123}^{II}$ and $B_{g,114}^{I}$ only depend on non-Gaussian initial conditions via small corrections to the matter power spectrum at mildly non-linear scales, \citep{TaruyaKoyamaMatsubara2008}, and have therefore no effect on the large-scale behavior. Following \citet{McDonald2006} and \citet{SmithScoccimarroSheth2007}, the terms $B_{g,123}^{II}$ and $B_{g,114}^I$ can be interpreted as providing a redefinition---or {\it renormalization}---of the bias parameters since they present the same functional dependence as $B_{g,112}^I$. Also, the not-renormalizable terms $B_{g,222}^I$ and $B_{g,123}^I$ behave, for equilateral configurations, like $P^2(k)$ as $k\rightarrow 0$, and we can expect them to be sub-leading with respect to a significant primordial contribution to the galaxy bispectrum.

Since we are mainly interested in the effects due to primordial non-Gaussianity at large scales, we will compute all perturbative corrections to the galaxy bispectrum up to order $\O(\d_0^5)$ in the linear {\it matter} density field, $\d_0$.  We will therefore ignore the 1-loop corrections to the matter bispectrum corresponding to the $B^{(6)}$ term as well as the non-linear bias terms corresponding to $B_{g,222}$, $B_{g,123}$ and $B_{g,114}$. Introducing the notation 
\beq
B_{g,ijk}(k_1,k_2,k_3)  \equiv  b_ib_jb_k\hat{B}_{g,ijk}(k_1,k_2,k_3),
\eeq
which simply factorizes the bias parameters from the quantities defined in Eqs.~(\ref{eq:Bg111}) to (\ref{eq:Bg114I}), we can rewrite the remaining contributions in terms of the expansions of Eq.s~(\ref{eq:Pexp}), (\ref{eq:Bexp}) and (\ref{eq:Texp}) for the matter power spectrum $P$, bispectrum $B$ and trispectrum $T$, respectively, to keep track of the perturbative order in terms of the linear matter density $\d_0$. We obtain, for the $B_{g,111}$ contribution simply the expansion of the matter bispectrum of Eq.~(\ref{eq:Bexp}), that is 
\beq
\hat{B}_{g,111}
 =    
B^{(3)}+B^{(4)}+B^{(5)}+\O(\d_0^6),
\eeq
where each component has been shown in the upper panels of Fig.~\ref{fig:bsmeq}. 
For the $B_{g,112}$ contribution we have instead two terms of order $\O(\d_0^4)$ and two of order $\O(\d_0^5)$, so that we can write
\beq
\hat{B}_{g,112}={B}_{g,112}^{(4)}+{B}_{g,112}^{(5)}+\O(\d_0^6)
\eeq
where ${B}_{g,112}^{(4)}={B}_{g,112}^{I,(4)}+{B}_{g,112}^{II,(4)}$ with
\bea
\hat{B}_{g,112}^{I,(4)}
& = &   
\tW(\kv_1,\kv_2)~P^{(2)}(k_1)P^{(2)}(k_2)+{\rm 2~perm.},
\\
\hat{B}_{g,112}^{II,(4)}
& = &   
\frac{1}{2}\intq~\tW(\qv,\kv_3\!-\!\qv)~T^{(4)}(\kv_1,\kv_2,\qv,\kv_3\!-\!\qv)+{\rm 2~perm.},
\eea
and where ${B}_{g,112}^{(5)}={B}_{g,112}^{I,(5)}+{B}_{g,112}^{II,(5)}$ with
\bea
\hat{B}_{g,112}^{I,(5)}
& = &   
\tW(\kv_1,\kv_2)~P^{(2)}(k_1)P^{(3)}(k_2)+{\rm 5~perm.},
\\
\hat{B}_{g,112}^{II,(5)}
& = &   
\frac{1}{2}\intq~\tW(\qv,\kv_3\!-\!\qv)~T^{(5)}(\kv_1,\kv_2,\qv,\kv_3\!-\!\qv)+{\rm 2~perm.}.
\label{eq:Bg112IIo5}
\eea
\begin{figure}[t]
\begin{center}
{\includegraphics[width=0.45\textwidth]{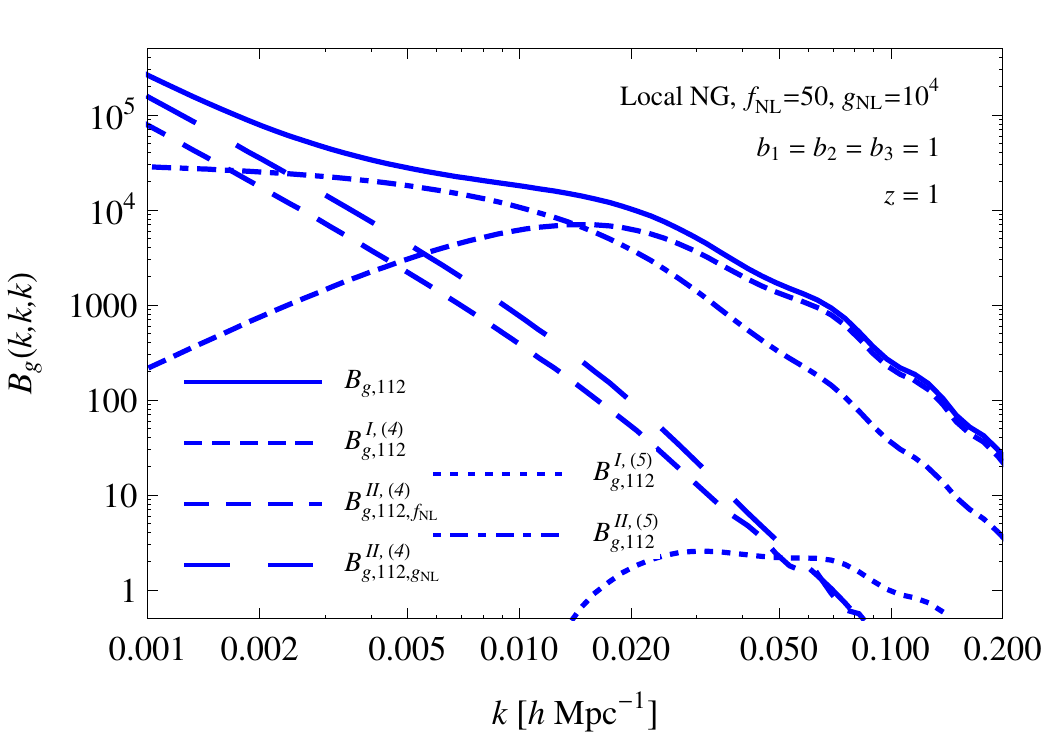}}
{\includegraphics[width=0.45\textwidth]{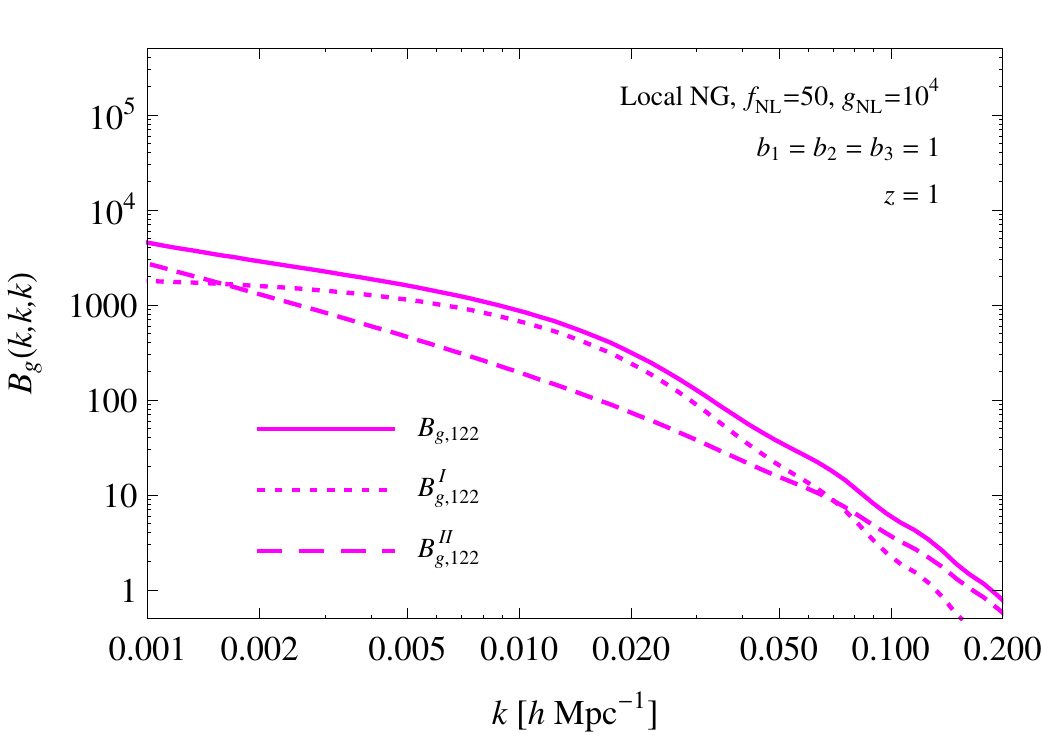}}
{\includegraphics[width=0.45\textwidth]{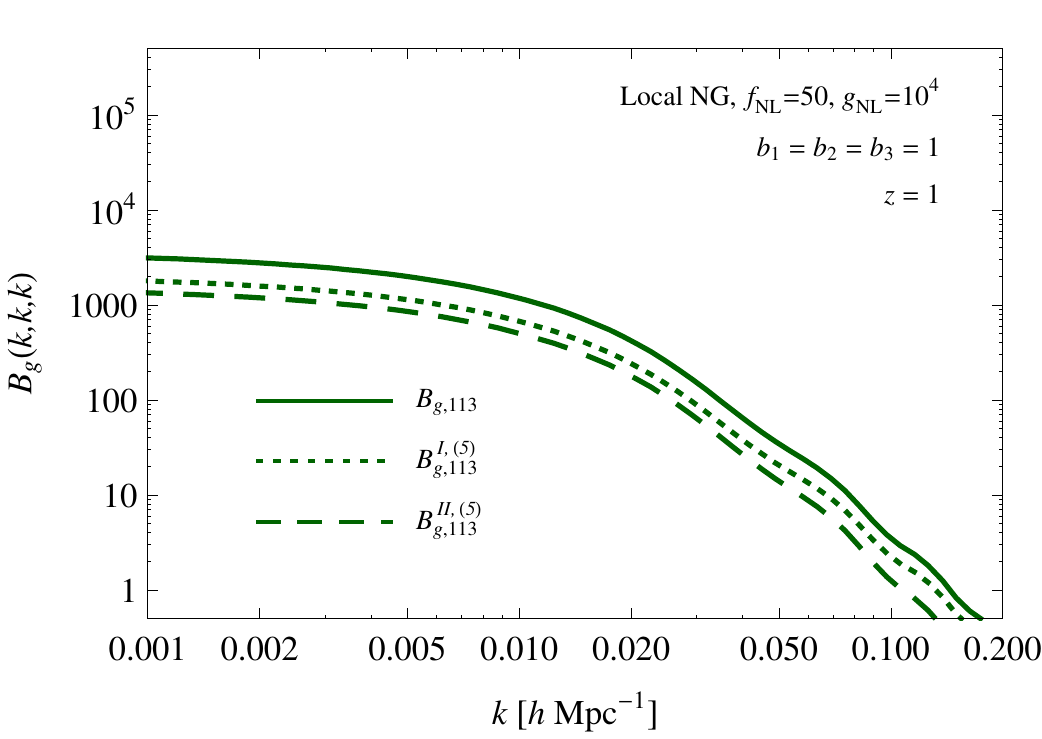}}
{\includegraphics[width=0.45\textwidth]{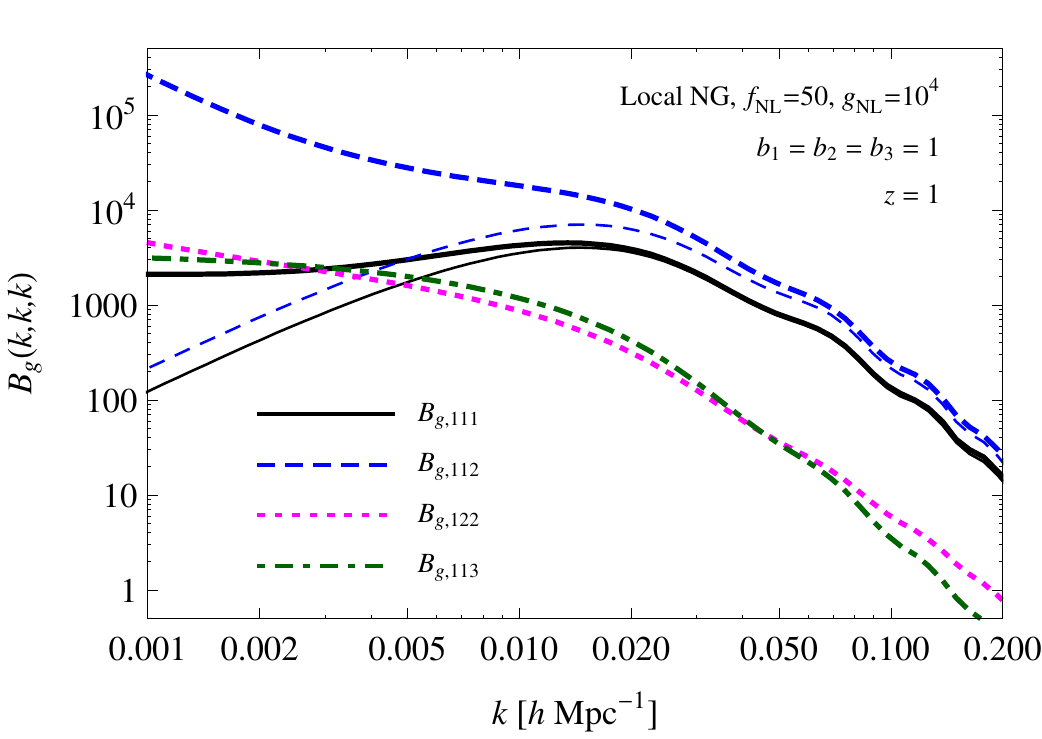}}
\caption{Contributions to the galaxy bispectrum with {\it local} non-Gaussian initial conditions assuming $\fNL=50$ and $\gNL=10^4$ and unitary bias parameters, $b_1=b_2=b_3=1$ at redshift $z=1$: $B_{g,112}$ ({\it upper left panel}), $B_{g,122}$ ({\it upper right panel}) and $B_{g,113}$ ({\it bottom left panel}). See the text for a detailed explanation of all terms. The bottom right panels shows a comparison of all different contributions, $B_{g,111}$ ({\it continuous black line}), $B_{g,112}$ ({\it dashed blue line}), $B_{g,122}$ ({\it dotted magenta line}) and $B_{g,113}$ ({\it dot-dashed green line}) with the thin lines corresponding to the values of  $B_{g,111}$ and  $B_{g,112}$ for Gaussian initial conditions. $B_{g,122}$ and  $B_{g,113}$, excluding corrections $\O(\d_0^6)$ are vanishing in the Gaussian case. Assumes a smoothing scale of $R=5\Mpc$.}
\label{fig:bsgLcz1}
\end{center}
\end{figure}
All this terms are show, for non-Gaussian initial conditions of the local kind and assuming $\fNL=50$ and $\gNL=10^4$, in the upper left panel of Fig.~\ref{fig:bsgLcz1}. Specifically, $B_{g,112}^{I,(4)}$ ({\it short-dashed line}) represent the usual, Gaussian tree-level prediction for galaxy bispectrum induced by quadratic non-linearities in the bias relation, while $B_{g,112}^{I,(5)}$ ({\it dotted line}) is a small correction to the same quantity due to the 1-loop contribution to the matter power spectrum $P^{(3)}$ for non-Gaussian initial conditions.

More interesting are the terms derived from $B_{g,112}^{II}$  depending on the matter trispectrum. For the local model we consider the non-vanishing initial trispectrum of Eq.~(\ref{eq:Tlc}) given by the sum of a term proportional to $\fNL^2$ and one proportional to $\gNL$. We plot these two components separately as $B_{g,112,\fNL}^{II,(4)}$ ({\it medium-dashed line}) and $B_{g,112,\gNL}^{II,(4)}$ ({\it long-dashed line}), with $B_{g,112}^{II,(4)}=B_{g,112,\fNL}^{II,(4)}+B_{g,112,\gNL}^{II,(4)}$. One can see that these terms provide the largest contribution to $B_{g,112}$ at large scales, with a particularly strong scale dependence. The large value $\gNL=10^4$ as been chosen here for illustration purposes only, in order to provide a contribution comparable in size to the one proportional to $\fNL^2$. In other words, for more natural values of $\gNL\sim$ few, we can expect this term to be negligible. Assuming the $\fNL^2$ term to be the leading contribution to the initial curvature trispectrum $T_\Phi$ given by Eq.~(\ref{eq:Tlc}) in the {\it local} non-Gaussian model, we have 
\bea
\hat{B}_{g,112,\fNL}^{II,(4)}(k_1,k_2,k_3)& \simeq & 2\fNL^2\intq~\tW(\qv,\kv_3\!-\!\qv)~M(k_1)M(k_2)M(\kv_3\!-\!\qv)M(q)
\nonumber\\
& & \times\left\{P_\Phi(k_1)P_\Phi(k_2)\left[P_\Phi(\kv_{13}\!-\!\qv)+P_\Phi(\kv_1\!+\!\qv)\right]+{\rm 5~perm.}\right\}+{\rm 2~perm.}
\eea
For equilateral configurations, in the large-scale limit this expression can be approximated as
\bea
\hat{B}_{g,112,\fNL}^{II,(4)}(k,k,k) & \stackrel{k\rightarrow 0}{\simeq}  &
36\fNL^2~M^2(k)~P_\Phi^2(k)\intq~W_R^2(q)~M^2(q)~P_\Phi(q)
\nonumber\\
& = & 36~\fNL^2~\sigma_R^2~M^2(k)~P_\Phi^2(k).
\eea
This asymptotic behavior can be compared to the one of the primordial contribution, $B^{(3)}=B_0$, given, for equilateral configurations, by $B_0(k,k,k)= 6~\fNL M^3(k)P_\Phi^2(k)$, so that 
\beq
\frac{\hat{B}_{g,112,\fNL}^{II,(4)}(k,k,k)}{B_0(k,k,k)}\stackrel{k\rightarrow 0}{\simeq} 6~\fNL\frac{\sigma_R^2(z)}{M(k,z)}\sim \frac{D(z)}{k^2},
\eeq
where we made explicit the dependence on redshift. We can therefore expect this component to be dominant at large scales, but its relevance with respect to the primordial one, decreasing with redshift. Notice that the redshift dependence is in part determined by the $\sigma_R^2(z)$ factor which might find a physical interpretation, as discussed at the end of the previous Section, when considered in its product with the quadratic bias parameter $b_2$. For our choice of the smoothing scale, $R=5\Mpc$, we have $\sigma_R(0)=1.07$ and we keep this factor in all evaluations shown in the figures.

We can also compare the asymptotic behavior with the Gaussian, tree-level prediction for the matter bispectrum, $B_{112}^I$, which is of the same order in PT. We have, again for equilateral configurations,
\beq
\frac{\hat{B}_{g,112,\fNL}^{II,(4)}(k,k,k)}{B_{112}^I(k,k,k)}\stackrel{k\rightarrow 0}{\simeq} \frac{7}{3}~\fNL^2\frac{\sigma_R^2(z)}{M^2(k,z)}\sim \frac{1}{k^4},
\eeq
where the ratio shows an even larger scale dependence, but is independent of redshift, as expected.

It is interesting to notice that the fifth-order contribution $B_{g,112}^{II,(5)}$ depending on the fifth-order matter trispectrum correction $T^{(5)}$ of Eq.~(\ref{eq:Bg112IIo5}), vanishing for Gaussian initial conditions is also equally important. We show this contribution as the dot-dashed line in the upper-left panel of Fig.~\ref{fig:bsgLcz1}, where one can see how it provides a significant contribution at intermediate scales between the trispectrum induced, 1-loop correction $B_{g,112}^{II,(4)}$ and the Gaussian tree-level term $B_{g,112}^{I,(4)}$. As discussed in Section~\ref{ssec:trisp}, this term arise due to the coupling between the initial bispectrum and quadratic non-linearities in PT. From the expression for $T_{1112}^I$ in Eq.~(\ref{eq:T1112}) one can write
\bea
\hat{B}_{g,112}^{II,(5)}(k_1,k_2,k_3) & = &  \intq~\tW(\qv,\kv_3\!-\!\qv)~
\nonumber\\
& & \times\left\{B_0(k_1,k_2,k_{12})\left[F_2(\kv_{12},\kv_3\!-\!\qv)P_0(\kv_3\!-\!\qv)+F_2(\kv_{12},\qv)P_0(q)\right]+{\rm 5~perm.}\right\}
\nonumber\\
& & +{\rm 2~perm.},
\eea
and in the large-scale limit, but for {\it generic} configurations and {\it generic} initial conditions one can show that
\beq
\hat{B}_{g,112}^{II,(5)}(k_1,k_2,k_3) \stackrel{k\rightarrow 0}{\simeq} \frac{34}{7}B_0(k_1,k_2,k_3)\sigma_R^2+\frac{160}{7}P_0(k_1)\intq W_R^2(q)B_0(k_2,q,q)+{\rm perm.},
\eeq
which shows how one of the permutation in Eq.~(\ref{eq:T1112}) provides a term proportional to initial matter bispectrum $B_0$, while the others an integration of the same. This means that we can expect this contribution to retain in part the shape-dependence of the initial bispectrum. Assuming local non-Gaussianity, for equilateral configurations, we can compare $B_{g,112}^{II,(5)}$ to $B_0$, obtaining the expected scale-independent behavior at large-scales
\beq
\frac{\hat{B}_{g,112}^{II,(5)}(k,k,k)}{B_0(k,k,k)}\stackrel{k\rightarrow 0}{\simeq} \frac{262}{21}\sigma_R^2(z)\sim D(z)^2,
\eeq
which shows as well how the relative weight of this contribution decreases with redshift. The ratio with the Gaussian term $B_{112}^{I}$ is given instead by
\beq
\frac{\hat{B}_{g,112}^{II,(5)}(k,k,k)}{B_{112}^I(k,k,k)}\stackrel{k\rightarrow 0}{\simeq} \frac{131}{3}\fNL\frac{\sigma_R^2(z)}{M(k,z)}\sim \frac{D(z)}{k^2}.
\eeq

\begin{figure}[t]
\begin{center}
{\includegraphics[width=0.45\textwidth]{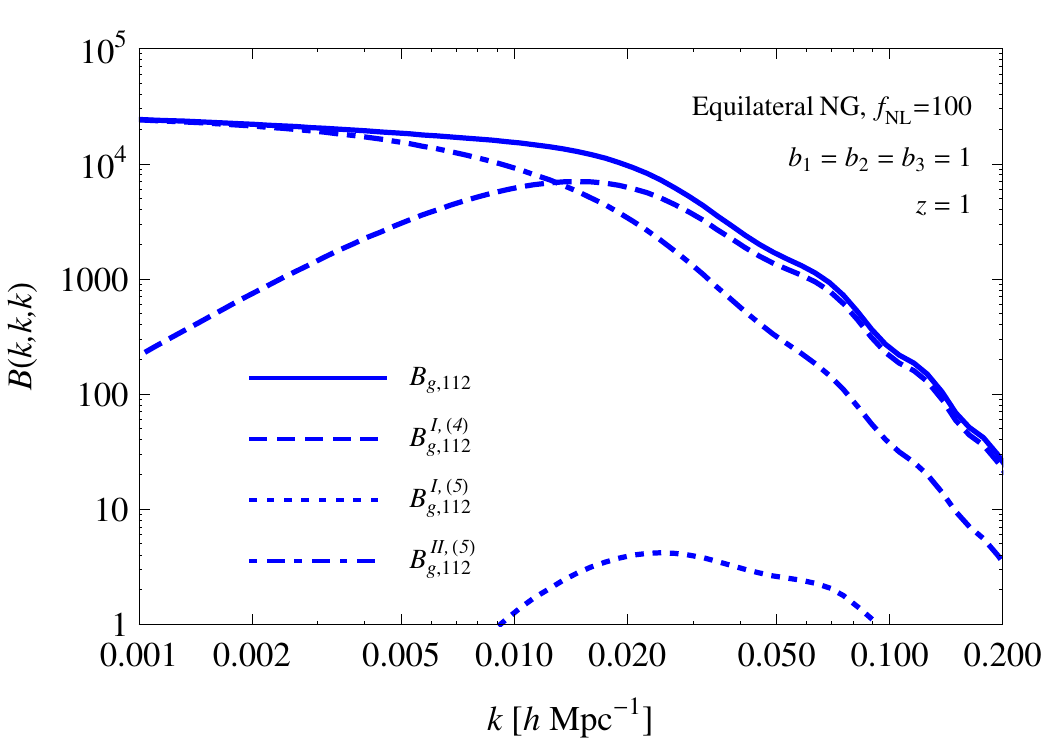}}
{\includegraphics[width=0.45\textwidth]{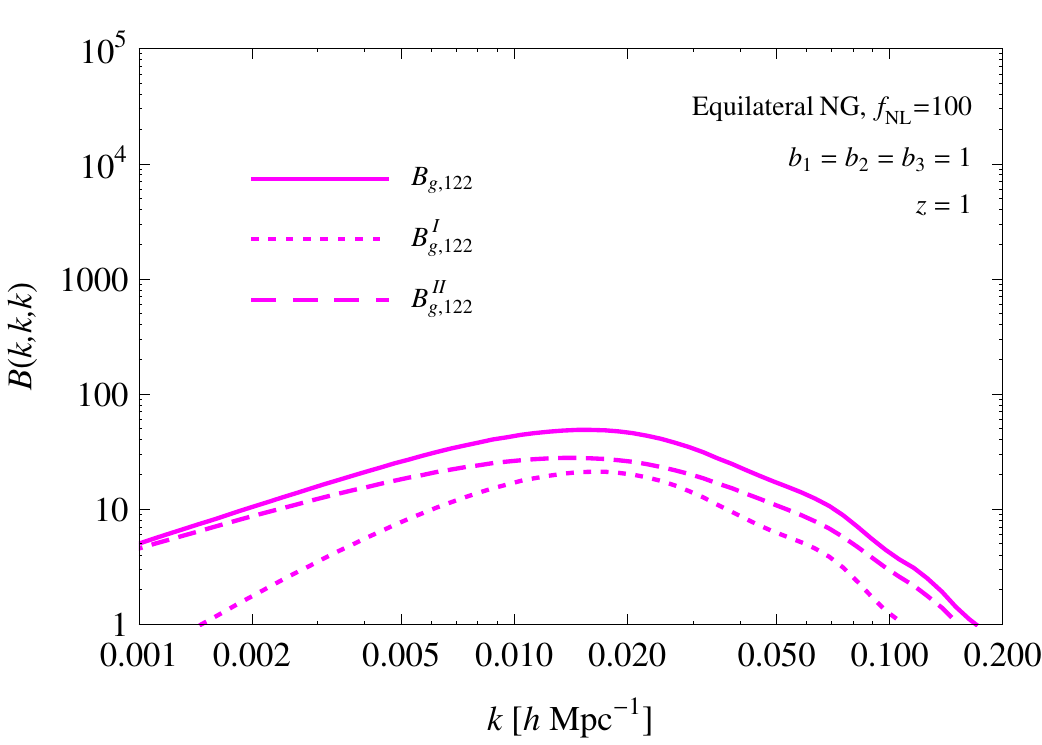}}
{\includegraphics[width=0.45\textwidth]{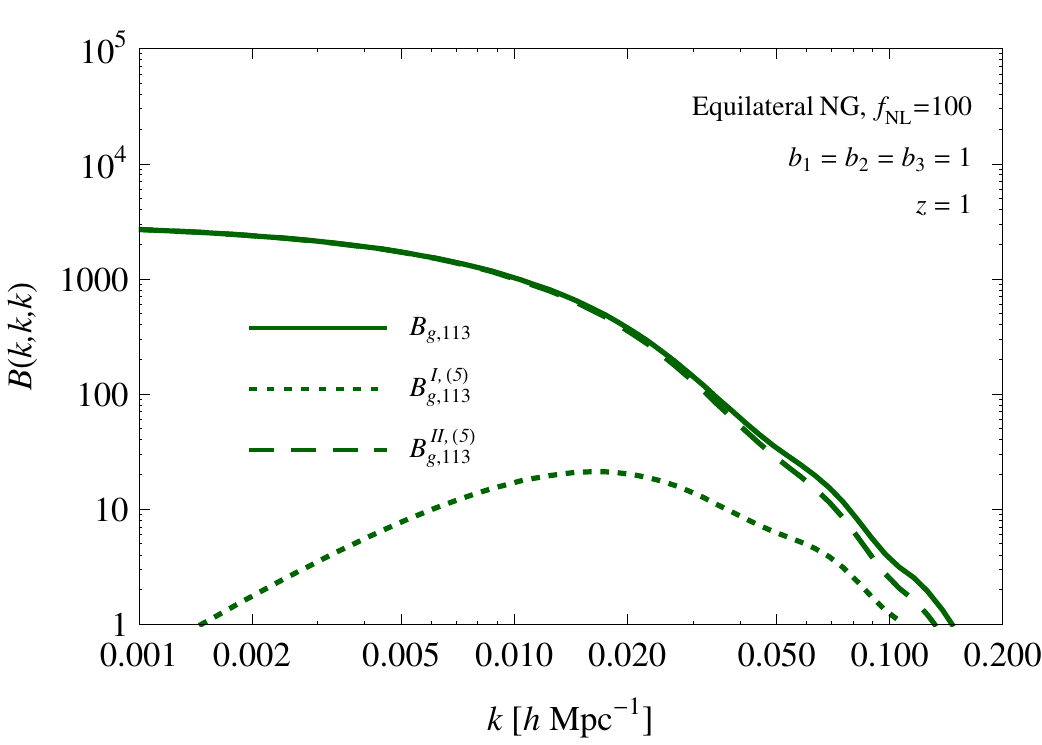}}
{\includegraphics[width=0.45\textwidth]{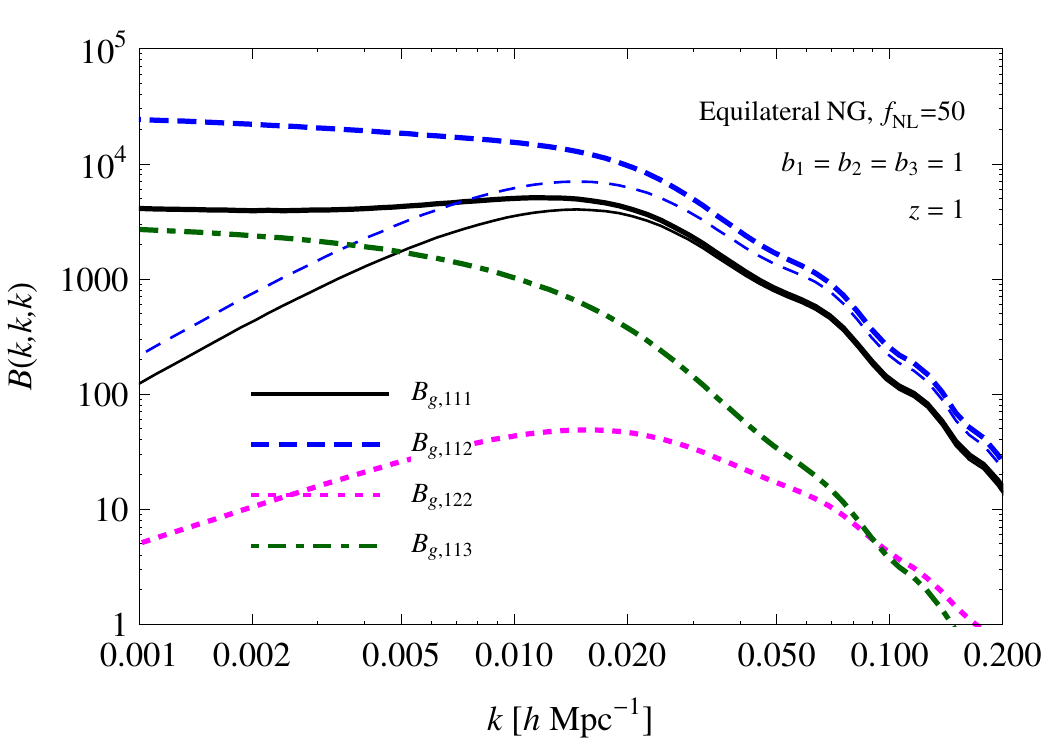}}
\caption{Same as Fig.~\ref{fig:bsgLcz1} but for equilateral non-Gaussianity with $\fNL=100$. In this case only a non-vanishing initial bispectrum is considered.}
\label{fig:bsgEqz1}
\end{center}
\end{figure}
In Fig.~\ref{fig:bsgEqz1} we show the same results as Fig.~\ref{fig:bsgLcz1} but for the {\it equilateral model}, assuming $\fNL=100$. Notice that in this case we consider only a non-vanishing initial bispectrum $B_0$, with no contributions from the initial trispectrum $T_0$ of the matter field, although they should be in principle taken into account for any realistic model. For this reason the $B_{g,112}^{II,(4)}$ terms are not present. It is remarkable, on the other hand, that the $B_{g,112}^{II,(5)}$ contribution due to the $T_{1112}^I$ term of Eq.~(\ref{eq:T1112}) in the expression for the matter trispectrum---dot-dashed line in the upper left panel of Fig.~\ref{fig:bsgEqz1}---{\it still provides a large correction at large-scales}. This behavior is specific to the galaxy bispectrum as it has been shown that lowest order correction to the galaxy power spectrum for acceptable levels of equilateral non-Gaussianity are negligible and probably unobservable, \citep{TaruyaKoyamaMatsubara2008,FedeliMoscardiniMatarrese2009}. 

We are left to consider the $B_{g,122}$ and $B_{g,113}$ terms contributing to our series for the galaxy bispectrum up to $\O(\d_0^5)$. These are shown in the upper right and bottom left panels, respectively, of Fig.~\ref{fig:bsgLcz1} for local non-Gaussian initial conditions with $\fNL=50$. Excluding corrections $\O(\d_0^6)$, both $B_{g,122}$ and $B_{g,113}$ are vanishing for Gaussian initial conditions. 

The $B_{g,122}$ consist in turn of the two terms, both of the fifth order in $\d_0$,
\bea
\hat{B}_{g,122}^{I,(5)}
& = &
\frac{1}{2} 
~\tW(\kv_1,\kv_3)~P^{(2)}(k_1)\intq~\tW(\qv,\kv_3\!-\!\qv)~B^{(3)}(k_3,q,|\kv_3\!-\!\qv|)+{\rm 5~perm.},
\\
\hat{B}_{g,122}^{II,(5)}
& = &
\intq~\tW(\qv,\kv_2\!-\!\qv)~\tW(\kv_1\!+\!\qv,\kv_2\!-\!\qv)B^{(3)}(k_1,q,|\kv_1\!+\!\qv|)~P^{(2)}(|\kv_2\!-\!\qv|)+{\rm 2~perm.}.
\eea
$B_{g,122}^I$ is represented in the upper right panel of Fig.~\ref{fig:bsgLcz1} for the local model and of Fig.~\ref{fig:bsgEqz1} for the equilateral model by the dotted magenta line, while $B_{g,122}^{II}$ by the dashed line. The continuous line is their sum. Their asymptotic behavior at large-scales, for equilateral configurations and assuming {\it local} initial conditions is given by
\bea
\hat{B}_{g,122}^{I,(5)}(k,k,k)
& \stackrel{k\rightarrow 0}{\simeq} &
12~\fNL P_0^2(k,z)\frac{\sigma_R^2(z)}{M(k,z)}\sim D(z)^5\quad{\rm [loc.~NG]},
\\
\hat{B}_{g,122}^{II,(5)}(k,k,k)
& \stackrel{k\rightarrow 0}{\simeq} &
12~\fNL P_0(k,z)\frac{1}{M(k,z)}\!\!\intq~ W_R^4(q)P^2_0(q)\sim \frac{D(z)^5}{k}\quad{\rm [loc.~NG]},
\eea
where we remark a peculiar form of scale-dependence in $B_{g,122}^{II,(5)}$, while for {\it equilateral} initial conditions, their contribution is suppressed at large scales
\bea
\hat{B}_{g,122}^{I,(5)}(k,k,k)
& \stackrel{k\rightarrow 0}{\simeq} &
12~\fNL M^3(k,z)P_\Phi^{4/3}(k)\!\!\intq~ W_R^2(q)M^2(q,z)P^{5/3}_\Phi(q)\sim D(z)^5 ~k^2\quad{\rm [eq.~NG]},
\\
\hat{B}_{g,122}^{II,(5)}(k,k,k)
& \stackrel{k\rightarrow 0}{\simeq} &
12~\fNL M(k,z)P_\Phi^{1/3}(k)\!\!\intq~ W_R^4(q)M^2(q,z)P_\Phi^{5/3}(q)P^2_0(q)\sim D(z)^5~k\quad{\rm [eq.~NG]}.
\eea

Finally, the $B_{g,113}$ contribution is made up by the two terms $B_{g,113}^{I,(5)}$ and $B_{g,113}^{I,(5)}$ shown, respectively by the dotted and dashed lines in the bottom left panel of Fig.~\ref{fig:bsgLcz1} for the local model and of Fig.~\ref{fig:bsgEqz1} for the equilateral model. It is clear from a comparison of Eq.~(\ref{eq:Bg113I}) with Eq.~(\ref{eq:Bg122I}) that term $\hat{B}_{g,113}^{I,(5)}$ is identical to $\hat{B}_{g,122}^{I,(5)}$ discussed above and it is therefore relevant at large-scales only for local non-Gaussianity. The term $\hat{B}_{g,113}^{II,(5)}$ instead is simply contributing to the overall multiplicative factor of the initial bispectrum, since
\beq
\hat{B}_{g,113}^{II,(5)}
 = 
\frac{3}{2}~ B_0(k_1,k_2,k_3)~\sigma_R^2(z).
\eeq
The term $B_{g,113}^I$ of Eq.~(\ref{eq:Bg113I}) therefore provides contributions equal to the $B_{g,111}$ term to all orders in the perturbative expansion of the matter bispectrum. However, in our choice to retain only corrections up to $\O(\d_0^5)$, the sole contribution proportional to the primordial bispectrum $B^{(3)}=B_0$ is present. This is consistent, in a way, with focusing our attention to the large-scale contributions due to primordial non-Gaussianity. 

We remind that, in this approximation, the only relevant initial correlators are the bispectrum $B_0$ and the trispectrum $T_0$ since we ignored as well the contribution of the initial five-point function to the trispectrum term $T^{(5)}$, in principle consistent with this expansion, but leading to an effective 2-loop correction with one integration in the expression for $T^{(5)}$ itself and a second one in the expression for $B_{g,112}^{II}$ above. Taking into account the identical contributions---up to a multiplicative factor---we could therefore rewrite Eq.~(\ref{eq:BgExp}) as follows
\beq
B_{g}
 \simeq 
b_1^3\left(1+\frac{3}{2}\frac{b_3}{b_1}\sigma_R^2\right)\hat{B}_{g,111}
+b_1^2b_2\left(\hat{B}_{g,112}^I+\hat{B}_{g,112}^{II}\right)
+b_1^3\left(\frac{b_2^2}{b_1^2}+\frac{b_3}{b_1}\right)\hat{B}_{g,122}^I+b_1b_2^2\hat{B}_{g,122}^{II},
\label{eq:BgRen}
\eeq
which shows that, in our approximation, only five different functional forms are present. Each multiplicative factor in the expression above should be interpreted in terms of the proper---{\it renormalized}---bias parameters to be compared with measurements in numerical simulations, \citep{McDonald2006,SmithScoccimarroSheth2007}.  
 
To explore the shape-dependence of the galaxy bispectrum, as in the case of the matter bispectrum, we can define a reduced galaxy bispectrum as
\beq
Q_g(k_1,k_2,k_3)\equiv\frac{B_g(k_1,k_2,k_3)}{\Sigma_g(k_1,k_2,k_3)},
\eeq
where
\beq
\Sigma_g(k_1,k_2,k_3)\equiv P_g(k_1)P_g(k_2)+{\rm cyc.}.
\eeq
Similarly to the bispectrum we introduce the notations
\bea
\Sigma_{g,ij,kl}(k_1,k_2,k_3)&\equiv& P_{g,ij}(k_1)P_{g,kl}(k_2)+{\rm perm.},
\\
& \equiv & b_ib_jb_lb_k\hat{\Sigma}_{g,ij,kl}(k_1,k_2,k_3),
\\
Q_{g,ijk}(k_1,k_2,k_3) & \equiv & \frac{b_ib_jb_k}{b_1^4}\hat{Q}_{g,ijk}(k_1,k_2,k_3),
\eea
to keep track of the non-linear order in the bias expansion and factorize the bias parameters. We can consider then the approximation
\beq
Q_{g}(k_1,k_2,k_3)\simeq\frac{1}{b_1}\hat{Q}_{g,111}+\frac{b_2}{b_1^2}\hat{Q}_{g,112}+\frac{b_2^2}{b_1^3}\hat{Q}_{g,122}+\frac{b_3}{b_1^2}\hat{Q}_{g,113}
\eeq
where we exclude higher-order contributions corresponding to corrections of order $\O(\d_0^2)$ and where
\bea
\hat{Q}_{g,111}
& = &
\frac{\hat{B}_{g,111}}{\hat{\Sigma}_{g,11,11}}=Q,
\\
\hat{Q}_{g,112}
& = &
\frac{\hat{B}_{g,112}-\hat{Q}_{g,111}\hat{\Sigma}_{g,11,12}}{\hat{\Sigma}_{g,11,11}},
\\
\hat{Q}_{g,122}
& = &
\frac{\hat{B}_{g,122}-\hat{Q}_{g,112}\hat{\Sigma}_{g,11,12}-\hat{Q}_{g,111}\left(\hat{\Sigma}_{g,11,22}+\hat{\Sigma}_{g,12,12}\right)}{\hat{\Sigma}_{g,11,11}},
\\
\hat{Q}_{g,113}
& = &
\frac{\hat{B}_{g,113}-\hat{Q}_{g,111}\hat{\Sigma}_{g,11,13}}{\hat{\Sigma}_{g,11,11}}.
\eea 

\begin{figure}[t!]
\begin{center}
{\includegraphics[width=0.45\textwidth]{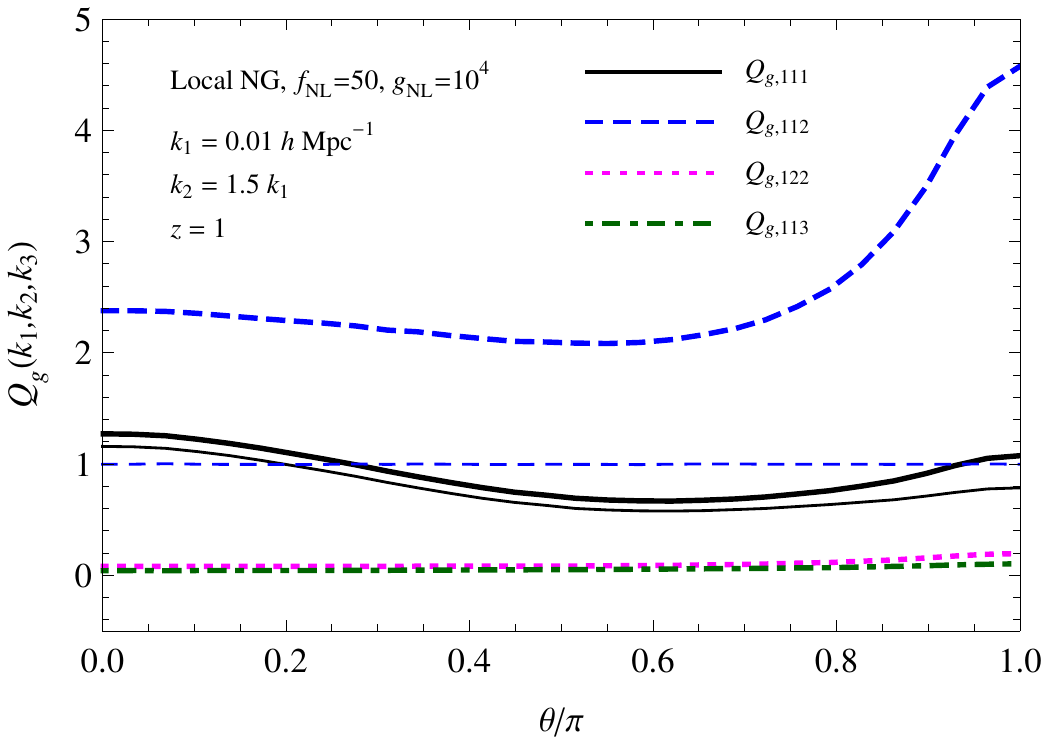}}
{\includegraphics[width=0.45\textwidth]{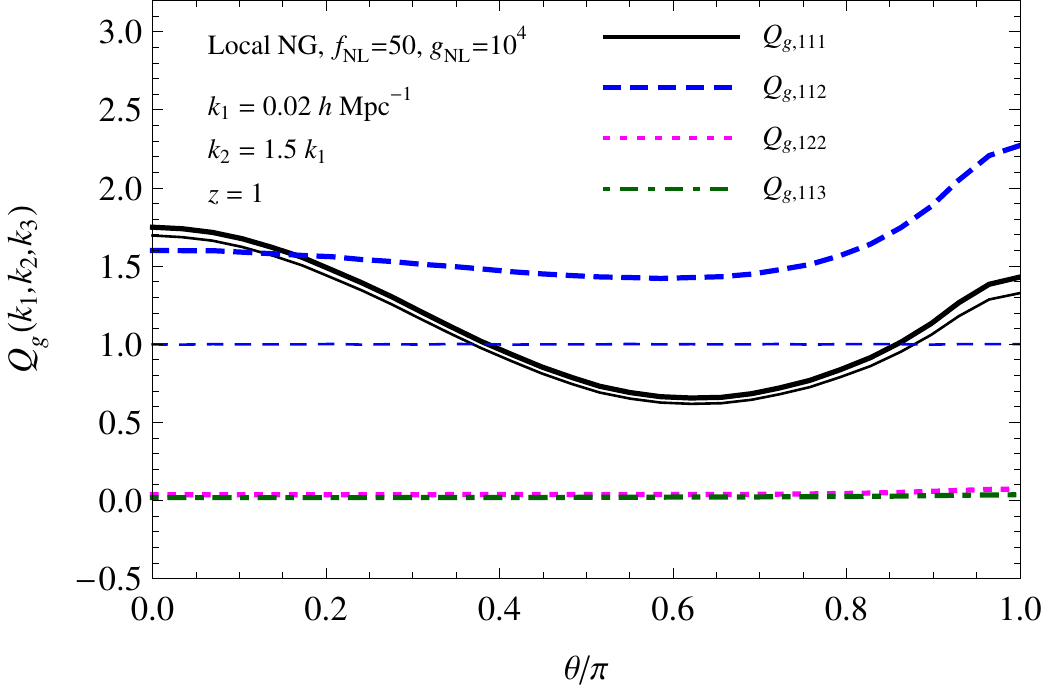}}
{\includegraphics[width=0.45\textwidth]{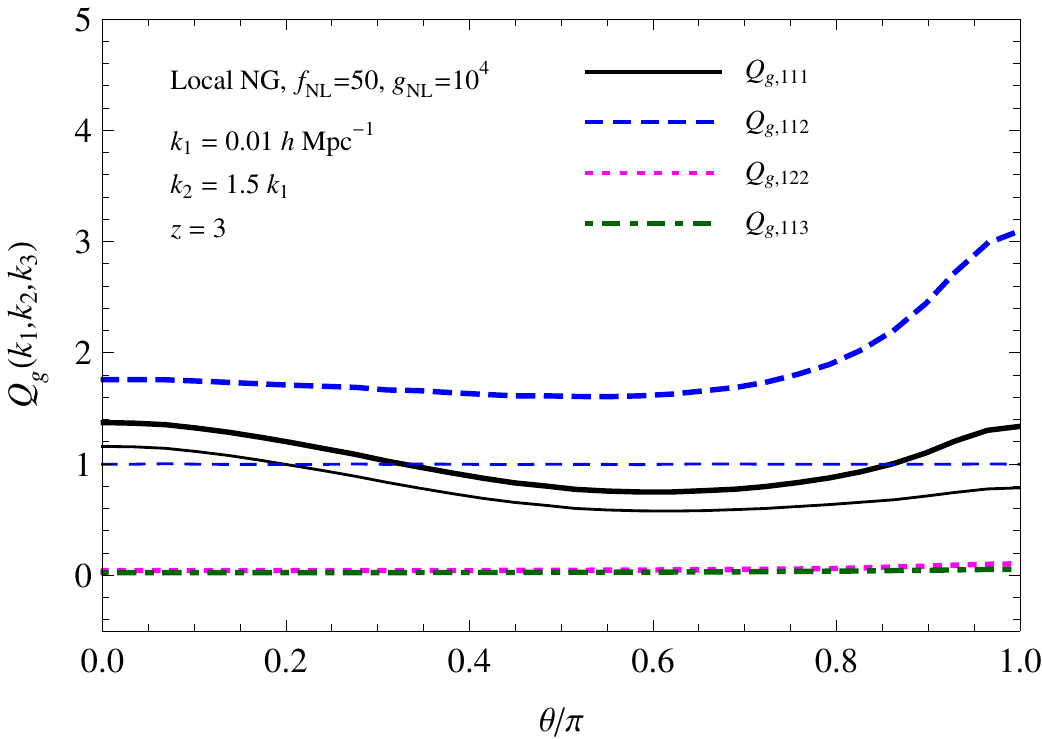}}
{\includegraphics[width=0.45\textwidth]{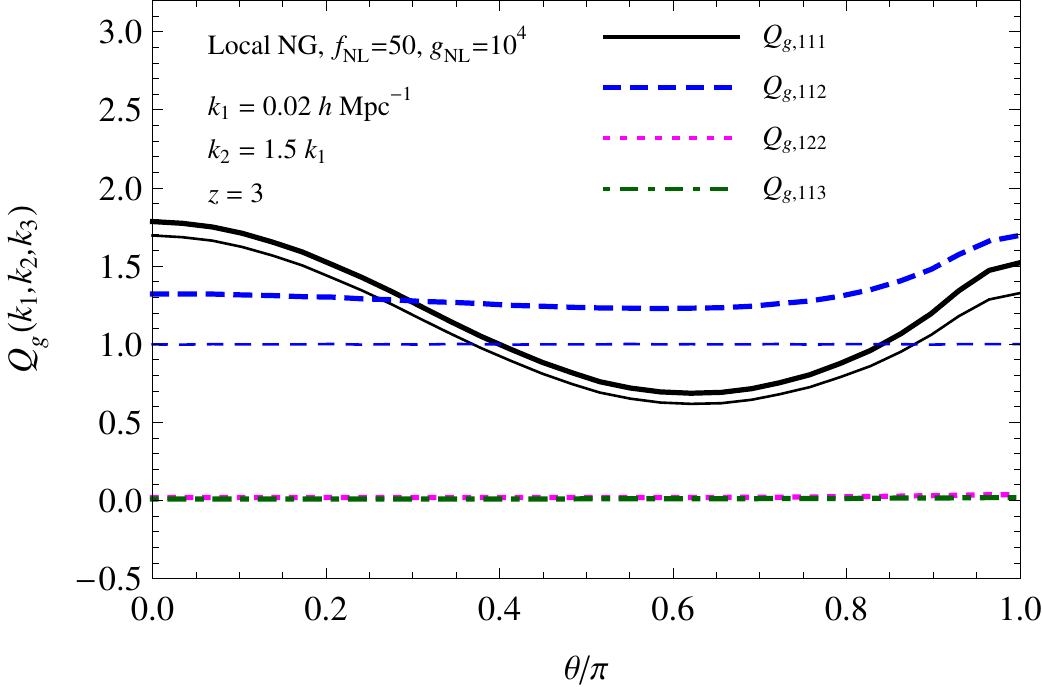}}
\caption{Galaxy bispectrum with local non-Gaussian initial conditions, assuming $\fNL=50$ and $\gNL=10^4$, as a function of the angle between the wavenumber $\kv_1$ and $\kv_2$  with $k_2=1.5~k_1$ for the different values $0.01$ ({\it left panels}) and $0.02\kMpc$ ({\it right panels}) and at redshift $z=1$ ({\it upper panels}) and $z=3$ ({\it lower panels}). Assumes unitary bias parameters $b_1=b_2=b_3=1$ and a smoothing scale of $R=5\Mpc$.}
\label{fig:qbsgLc}
\end{center}
\end{figure}
\begin{figure}[t!]
\begin{center}
{\includegraphics[width=0.45\textwidth]{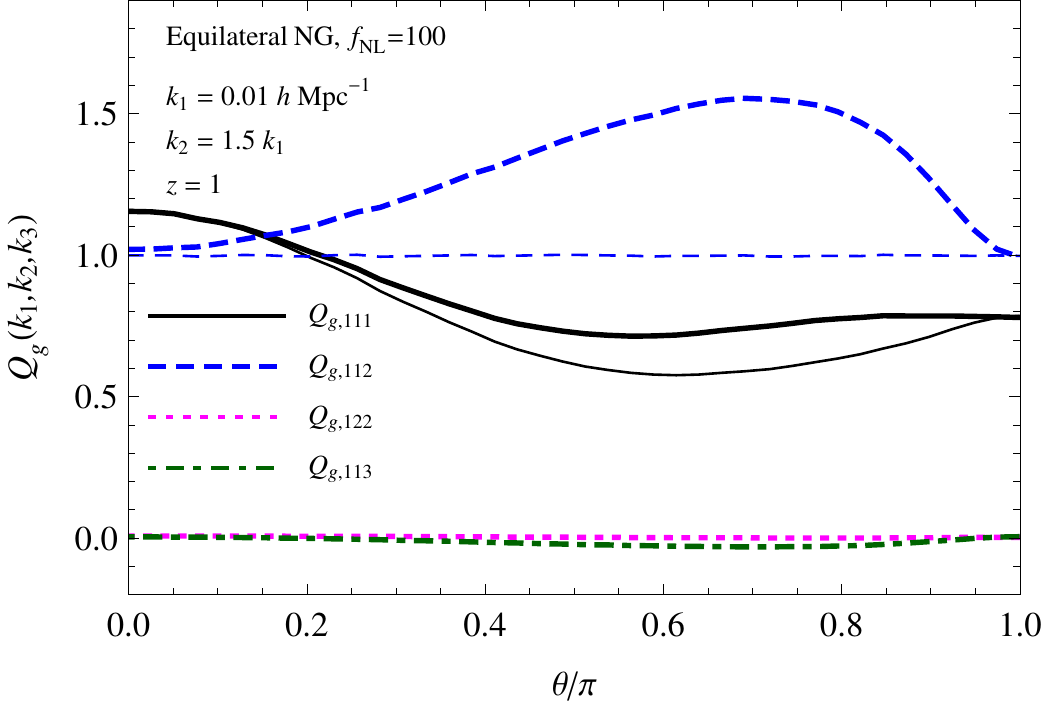}}
{\includegraphics[width=0.45\textwidth]{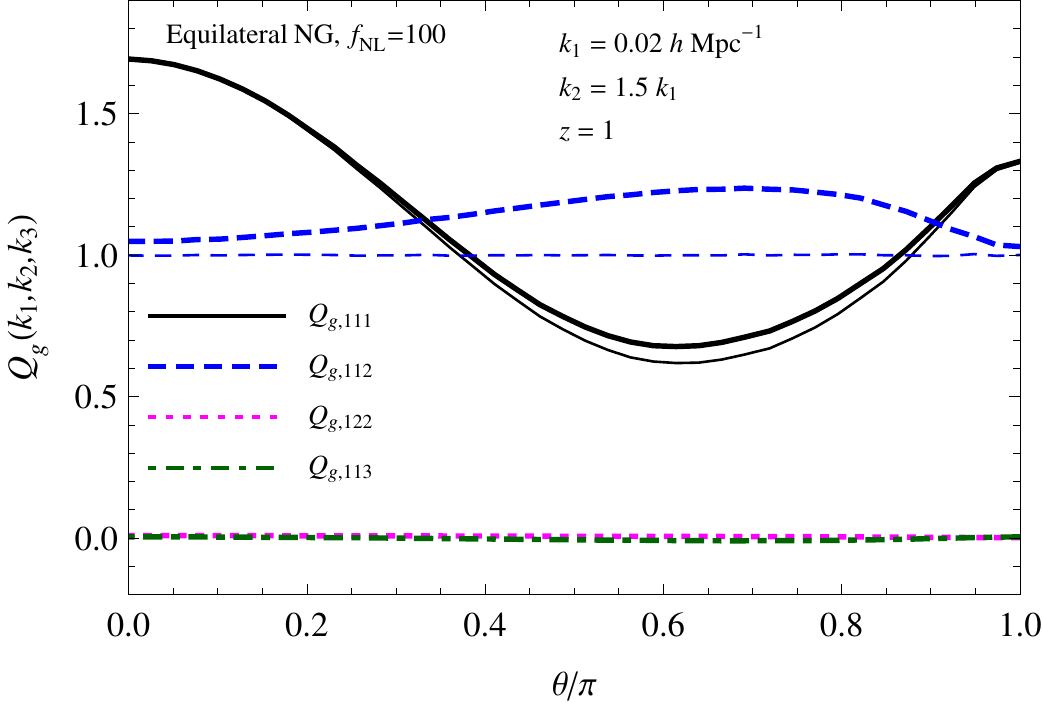}}
{\includegraphics[width=0.45\textwidth]{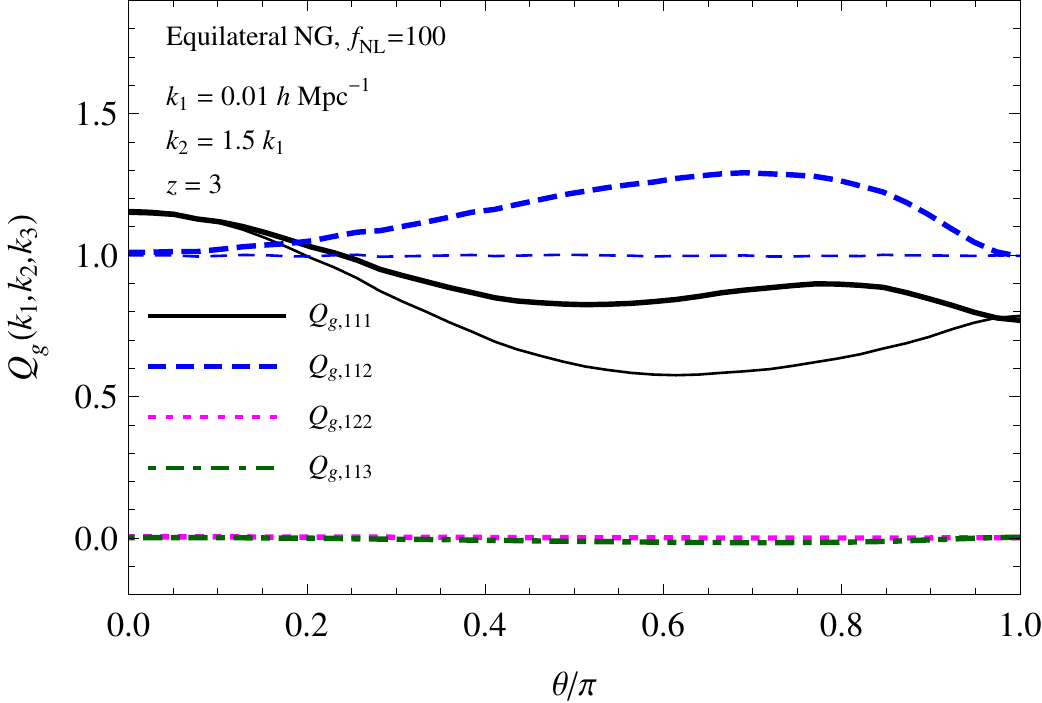}}
{\includegraphics[width=0.45\textwidth]{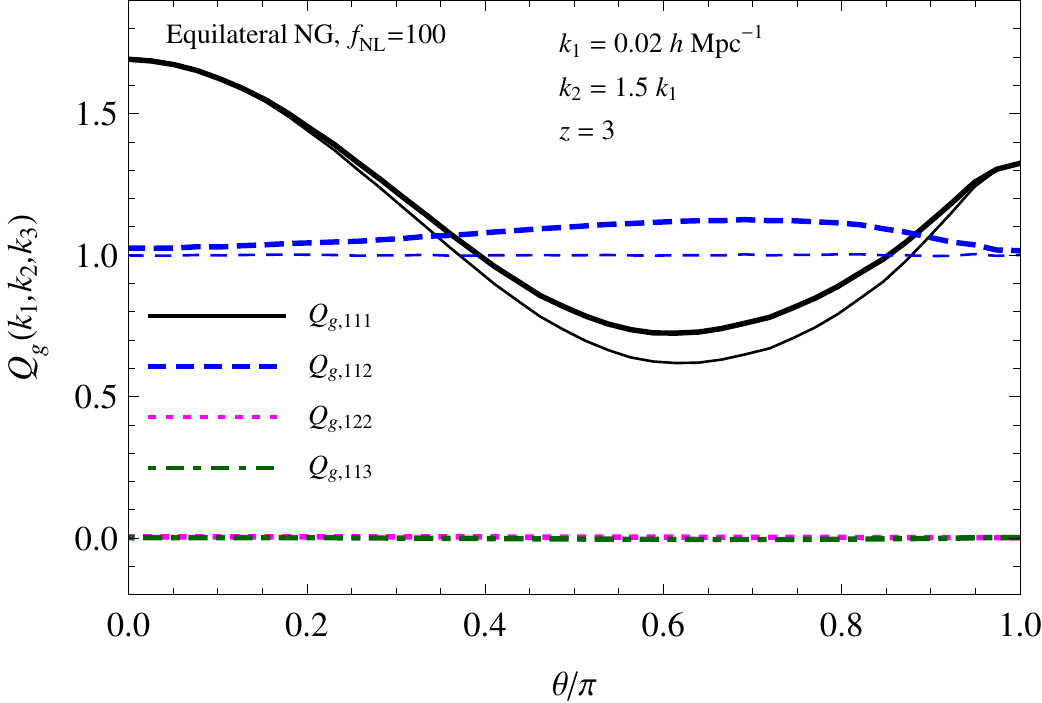}}
\caption{Same as Fig.~\ref{fig:qbsgLc} but for equilateral non-Gaussianity.}
\label{fig:qbsgEq}
\end{center}
\end{figure}

 
In Fig.~\ref{fig:qbsgLc} we plot the different contributions to the reduced galaxy bispectrum with {\it local} non-Gaussian initial conditions, assuming $b_1=b_2=b_3=1$, as a function of the angle between the vectors $\kv_1$ and $\kv_2$ with $k_2=1.5\, k_1$ for two different values of the magnitude $k_1$ and at redshifts $z=1$ ({\it left panels}) and $z=3$ ({\it right panels}). Fig.~\ref{fig:qbsgEq} shows the same results for {\it equilateral} non-Gaussian initial conditions. Thick lines represent the sum of all contributions to each term $Q_{g,ijk}$ while the thin lines for $Q_{g,111}$ and $Q_{g,112}$ correspond to their values for Gaussian initial conditions. 
 
In these plots it is particularly evident the difference between the two non-Gaussian models, despite the fact that we did not choose a set of triangular configurations including particularly squeezed triangles since, at most, for $\theta=\pi$ we have $k_2=1.5~k_1=3~k_3$. Still, one can observe for the local model in  Fig.~\ref{fig:qbsgLc} a significant increase in the non-Gaussian signal as $\theta\rightarrow \pi$, both in the correction to the matter bispectrum contained in $Q_{g,111}$ and even more so in the lowest order correction due to non-linear, quadratic bias $Q_{g,112}$. Indeed, as expected from the plots of the equilateral configurations of Fig.~\ref{fig:bsgLcz1}, $Q_{g,112}$ is responsible for the largest non-Gaussian signature. Clearly, as mentioned before, choosing a set of triangles with $k_1\sim k_2$ would lead to an even greater effect, and, in the limiting case $k_1=k_2$, the two contributions diverge as $\theta\rightarrow \pi$. Such extreme triangles are, however, difficult to measure in real surveys and we limit ourselves to provide an idea of the non-Gaussian corrections to the galaxy bispectrum for quite generic triangular configurations. The signature of an equilateral primordial non-Gaussianity is, on the other hand, evident, in the plots of Fig.~\ref{fig:qbsgEq}, where the largest contribution is expected for $\theta\sim0.7\,\pi$, the closest to an equilateral configuration. Again, regardless of the effect due to the initial trispectrum which we do not include here, the largest effect is on the $Q_{g,112}$ term, which departs significantly from the constant predicted by the Gaussian, tree-level expression, due essentially to the $B_{g,112}^{II,(5)}$ contribution discussed above, that is
\beq
Q_{g,112}=\frac{b_2}{b_1^2}\quad {\rm for }~~\fNL=0
\eeq
which corresponds to the value 1 in the plots where we set $b_1=b_2=b_3=1$. Indeed we find that such correction, for a level of non-Gaussianity well within current constraints such as $\fNL=50$, can be a factor of a few larger than the Gaussian prediction already at  $k\simeq 0.01\kMpc$ and at redshift $z=1$, that is, at scales nearly accessible by current surveys. The overall impact on the galaxy bispectrum depends, of course, on the relative size of non-linearities in the galaxy bias relation, or, in simple terms, on the ratio $b_2 /b_1$. Such ratio, assuming the halo model predictions for Gaussian initial conditions, is expected to be small, and in fact negative, for galaxy samples like the main samples of the Sloan Digital Sky Surveys (SDSS) or of the 2dF Galaxy Redshift Survey, but it is also expected to increase with an increasing value of the linear bias, $b_1$, and be therefore particular significant for populations like the Luminous Red Galaxies sample of the SDSS,  \citep{VerdeEtal2002,GaztanagaEtal2005,NishimichiEtal2007}.

We can also notice that, while the non-Gaussian effect on the $Q_{g,111}$, mostly due to the primordial component $B_0$ to the matter bispectrum, increases with redshift, the effects due to the combination of non-linear bias with non-Gaussian initial conditions instead decrease with redshift. This observation is in contrast with the notion that high-redshift surveys will provide the best constraints on non-Gaussianity, based in turn on analyses which took into account only the detectability of the primordial component, \citep{SefusattiKomatsu2007}. Quantitive predictions for the expected constraints on non-Gaussian parameters from measurements of the galaxy bispectrum in upcoming surveys will however require further work and we leave them entirely for future publications.

\section{On the high-peaks three-point function}
\label{sec:peaks}
 
In this Section we comment on the differences between our results and the similar ones derived by \citet{JeongKomatsu2009}, following a different approach that extends the study of the high-peak two-point correlation function and power spectrum of \citet{MatarreseVerde2008} to the high-peak bispectrum.  
 
As mention already at the end of Section~\ref{sec:GalaxyPS}, \citet{MatarreseVerde2008} derived a correction to the peak two-point correlation function $\xi_{p,M}$ in terms of the three-point correlation function, 
\beq\label{eq:2pcfMV}
\xi_{p}(\xv_1,\xv_2)\simeq\frac{\nu^2}{\sigma_R^2}\xi_{0,R}(\xv_1,\xv_2)+\frac{\nu^3}{\sigma_R^3}\zeta_{0,R}(\xv_1,\xv_2,\xv_2),
\eeq
where $\xi_{0,R}$ and $\zeta_{0,R}$ correspond to the two- and three-point function of the {\it initial} density perturbations, filtered on a scale $R$ and where $\nu\equiv\d_c/\sigma_R$ with $\sigma_R$ being the r.m.s. of the smoothed linear fluctuations.

From our assumption of local bias, Eq.~(\ref{eq:LocalBias}), we can write a similar expression for the galaxy two-point function,
\beq\label{eq:2pcfLcBias}
\xi_{g}(\xv_1,\xv_2)\simeq b_1^2\xi_{R}(\xv_1,\xv_2)+b_1b_2\zeta_{R}(\xv_1,\xv_2,\xv_2),
\eeq
where $\xi_R$ and $\zeta_R$ are the filtered and {\it evolved} matter two- and three-point functions evaluated at the redshift of interest. Both expressions give rise to similar corrections to the power spectrum, particularly large when the initial three-point function assumes large values for squeezed triangular configurations, as it is the case for local non-Gaussianity. We can notice, in fact, that the three-point functions that appear in Eq.~(\ref{eq:2pcfMV}) and (\ref{eq:2pcfLcBias}) are evaluated for collapsed triangles. The second expression, however, includes all higher-order corrections to the matter bispectrum in PT as described by Eq.~(\ref{eq:Bexp}). In addition, a comparison between the two leads to the improper identifications $b_1\sim \nu/\sigma_R$ and $b_2\sim \nu^2/\sigma_R^2$ for the bias parameters in the high-peaks limit, $\nu\gg 1$. In fact, these values correspond to the high-threshold limit for the Eulerian expressions of the halo bias parameters derived in the framework of the ellipsoidal collapse, \citep{MoJingWhite1997,ScoccimarroEtal2001A}, but represent Lagrangian quantities in Eq.~(\ref{eq:2pcfMV}).

\citet{MatarreseLucchinBonometto1986}, on which the results of \citet{MatarreseVerde2008} are based, derived as well an expression for higher-order correlators and, in particular, for the three-point function we have,
\bea\label{eq:MLB}
\zeta_p(\xv_1,\xv_2,\xv_3)&=&F(\xv_1,\xv_2,\xv_3)\left[\xi_p(\xv_1,\xv_2)\xi_p(\xv_1,\xv_3)\xi_p(\xv_2,\xv_3)\right.
\nonumber\\
& &
\left.+\xi_p(\xv_1,\xv_2)\xi_p(\xv_2,\xv_3)+{\rm 2~perm.}+\xi_p(\xv_1,\xv_2)+{\rm 2~perm.}+1\right]
\nonumber\\
& &-\xi_p(\xv_1,\xv_2)+{\rm 2~perm.}-1,
\eea
where
\beq
F(\xv_1,\xv_2,\xv_3)=\exp\left\{\sum_{n=3}^{\infty}\sum_{j=1}^{n-2}\sum_{k=1}^{n-j-1}\frac{(\nu/\sigma_R)^n}{j!k!(n-j-k)!}\xi_{0,R,[j;k;n-j-k]}^{(n)}\right\},
\eeq
with $\xi_{0,R,[j;k;n-j-k]}^{(n)}$ representing the smoothed linear matter $n$-point function with the $n$ arguments given by $j$ times $\xv_1$, $k$ times $\xv_2$ and $(n-j-k)$ times $\xv_3$. Expanding the exponential for small values of the correlators to include corrections up to the matter four-point function, $\xi^{(4)}$, one obtains
\beq
F(\xv_1,\xv_2,\xv_3) \simeq  1+\frac{\nu^3}{\sigma^3_R}\zeta_{0,R}(\xv_1,\xv_2,\xv_3)+
\frac{3}{2}\frac{\nu^4}{\sigma^4_R}\xi_{0,R}^{(4)}(\xv_1,\xv_2,\xv_3,\xv_3).
\eeq
Retaining this order of corrections, proportional to the forth power in the linear density field, \citet{JeongKomatsu2009} derive from Eq.~(\ref{eq:MLB}), the expression for the bispectrum of high-peaks given by
\bea\label{eq:zetap}
\zeta_p(\xv_1,\xv_2,\xv_3) & \simeq & \frac{\nu^3}{\sigma^3_R}\zeta_{0,R}(\xv_1,\xv_2,\xv_3)+
\frac{1}{2}\frac{\nu^4}{\sigma^4_R}\xi_{0,R}^{(4)}(\xv_1,\xv_2,\xv_3,\xv_3)+{\rm 2~perm.}\nonumber\\ & &
+\frac{\nu^4}{\sigma_R^4}\xi_{0,R}(\xv_1,\xv_3)\xi_{0,R}(\xv_2,\xv_3)+{\rm 2~perm.}
\eea

Again, we can compare this expression with the similar one that can be obtained from the local bias prescription of Eq.~(\ref{eq:LocalBias}), that is
\beq
\la\d_g(\xv_1)\d_g(\xv_2)\d_g(\xv_3)\ra\simeq b_1^3\la\d(\xv_1)\d(\xv_2)\d(\xv_3)\ra+\frac{1}{2}b_1^2b_2\la\d(\xv_1)\d(\xv_2)\d^2(\xv_3)\ra+{\rm 2~perm.}
\eeq
where the four-point function $\la\d(\xv_1)\d(\xv_2)\d^2(\xv_3)\ra$ is given by a connected and a {\it non}-connected component corresponding, respectively, to the second and third terms on the r.h.s. of Eq.~(\ref{eq:zetap}), so that the galaxy three-point function is given by
\bea\label{eq:zetag}
\zeta_g(\xv_1,\xv_2,\xv_3) & \simeq & b_1^3\zeta_{R}(\xv_1,\xv_2,\xv_3)+
\frac{1}{2}b_1^2b_2\xi_{R}^{(4)}(\xv_1,\xv_2,\xv_3,\xv_3)+{\rm 2~perm.}\nonumber\\ & &
+b_1^2b_2\xi_{R}(\xv_1,\xv_3)\xi_{R}(\xv_2,\xv_3)+{\rm 2~perm.}
\eea
In the approach of Section~\ref{sec:galaxyPT}, however, the correlators $\xi_R$, $\zeta_R$ and $\xi_R^{(4)}$ (or their {\it unsmoothed} versions) are fully evolved, while the corresponding quantities $\xi_{0,R}$, $\zeta_{0,R}$ and $\xi_{0,R}^{(4)}$ of Eq.~(\ref{eq:zetap}) are their initial counterparts. This implies that Eq.~(\ref{eq:zetap}) does not include non-Gaussianities due to gravitational instability but only those determined by non-linear bias and the initial conditions. \citet{JeongKomatsu2009}, however, assume that the {\it initial} correlators correspond to relatively late-time ($z\sim 10$) quantitities so that, in their expression for the halo bispectrum, based on Eq.~(\ref{eq:zetap}), the matter correlators do include contributions due to the growth of structures. Their result is therefore formally equivalent to the one of Eq.~(\ref{eq:zetag}), leading to the same non-Gaussian corrections, particularly since, under their assumption, the predictions of Eq.~(\ref{eq:zetap}) for the bias factors loose their validity and are replaced by parameters to be fitted by comparison with observations.
\section{Conclusions}
\label{sec:conclusions}

In this work we studied several aspects of the effects of non-Gaussian initial conditions, in the framework of Eulerian perturbation theory, both on the {\it matter} as on the {\it galaxy} bispectrum under the assumption of local bias. 

In Section~\ref{sec:matterPT}, we have computed perturbative 1-loop corrections to the matter bispectrum in presence of a primordial non-Gaussian component corresponding to two common phenomenological models resulting in a non-vanishing initial matter bispectrum, and, for the local model of non-Gaussianity, also in an initial matter trispectrum. We have shown that, as it is the case for the power spectrum, the effects of such corrections, given current constraints from CMB observations on the amplitude of non-Gaussianities both of the local and equilateral kind, at mildly non-linear scales are of the order of a few percent of the leading contribution to the matter bispectrum due to gravitational instability. 

In Section~\ref{sec:galaxyPT} we considered then a perturbative expansion from a local bias prescription and the corresponding corrections to the galaxy bispectrum, and identified several large contributions enhanced by non-linear bias that might exceed the direct effect due to the initial component. In particular we have shown that, for local primordial non-Gaussianity, a {\it significant non-Gaussian effect is expected due to the non-vanishing matter trispectrum}, both by its primordial component in the local model, depending on the initial trispectrum, at very large scales, thus confirming recent findings by \citet{JeongKomatsu2009}, as well as by the next-to-leading order correction in perturbation theory, depending instead on the initial bispectrum, which appears to be {\it larger at smaller scales}. In addition this last contribution is significant also for {\it equilateral} non-Gaussianities, whose effects on the galaxy power spectrum have been shown to be negligible, \citep{TaruyaKoyamaMatsubara2008}.  

Specifically, for local non-Gaussianities, these two contributions present a large-scale behavior characterized by extra factors of $1/k^4$ and $1/k^2$, respectively, when compared to the predictions for the galaxy bispectrum for Gaussian initial conditions. This implies that, for significant quadratic non-linearities in the galaxy bias, they are expected to represent the leading contribution to the galaxy bispectrum on slightly different ranges of scales, with the component induced by the initial trispectrum dominant at the largest scales and the one induced by the initial bispectrum dominant on intermediate scales of the order of $k\sim 0.005\kMpc$, while Gaussian terms represent the leading component on smaller scales. 

For equilateral non-Gaussianities we assumed, for simplicity, a vanishing initial trispectrum. However, the contribution to the matter trispectrum due to the initial bispectrum alone, still leads to a significant effect on the galaxy bispectrum. This appears to be a generic effect of primordial non-Gaussianity, regardless of the model, while no analogous effect is present for the galaxy power spectrum.  

We can therefore expect the galaxy bispectrum to be a probe of primordial non-Gaussianity sensitive to the initial bispectrum {\it as well as} to the initial trispectrum. This is not surprising given the effects that the initial bispectrum has on the galaxy power spectrum. However, it is particularly interesting because, despite the fact that the simple local model of Eq.~(\ref{eq:phiNG}) predicts a trivial relation between the amplitudes of the primordial bispectrum and trispectrum, with the latter given just by $\fNL^2$, in general we can expect the determinations of these two quantities to provide distinct constraints on the parameters of the inflationary model. Moreover, the sensitivity of the galaxy bispectrum on primordial non-Gaussianity does not seems to be restricted to the local model, as it appears to be the case for the galaxy power spectrum.

A comparison of these results for the galaxy,  and halo, bispectrum with numerical simulations is, at this point, necessary, to obtain reliable quantitative predictions to be used in the analysis of large-scale observations. Furthermore, the study of the effect of non-Gaussian initial conditions on the halo bispectrum can significantly improve our theoretical understanding of the effect on the halo power spectrum itself, as we can well expect it to point-out unavoidable shortcomings and inaccuracies of the different approaches proposed so far in the literature, all assuming somehow different approximations. 

\citet{ScoccimarroSefusattiZaldarriaga2004} and \citet{SefusattiKomatsu2007} have shown that constraints on non-Gaussianity of both local and equilateral kind from measurements of the galaxy bispectrum in future redshift surveys are comparable and even better than CMB constraints. However, these works assumed the tree-level prediction in perturbation theory for the galaxy bispectrum and neglected contributions depending on the matter trispectrum. An update of these results is clearly necessary, in light of the results of this paper as well as those of \citet{JeongKomatsu2009} and we will leave it for future work. If the qualitative results presented here were to be confirmed in numerical simulations, it is reasonable to assume that such predictions will significantly improve. In particular, a combined analysis of power spectrum and bispectrum of biased populations at large-scales, when an accurate description of non-linearities in the bias relation is achieved, might represent the best probe of a departure from Gaussian initial conditions, the latter defined by a non-vanishing primordial bispectrum {\it as well as} a non-vanishing primordial trispectrum and possibly higher-order correlators.

\acknowledgments

I thank Francis Bernardeau and Rom\'an Scoccimarro for useful discussions and a careful reading of the manuscript. I am  particularly grateful to the staff of Nordita for its kind support and hospitality even past my visiting term. I acknowledge support by the French Agence National de la Recherche under grant BLAN07-1-212615.

\bibliography{Bibliography}

\end{document}